\documentclass[aps,prl,twocolumn,showpacs,superscriptaddress,floatfix]{revtex4-1}
\usepackage{amsmath,amssymb}
\usepackage{graphicx}
\usepackage{epsfig}
\usepackage{color}
\usepackage{rotating}
\usepackage{bm}
\usepackage{setspace}
\begin{document}
%Title of paper
\title{Percolative Superconductivity in Electron-Doped Sr$_{1-x}$Eu$_{x}$CuO$_{2+y}$ Films}
\author{Xue-Qing Yu}
\author{Hang Yan}
\author{Li-Xuan Wei}
\author{Ze-Xian Deng}
\author{Yan-Ling Xiong}
\author{Jia-Qi Fan}
\affiliation{State Key Laboratory of Low-Dimensional Quantum Physics, Department of Physics, Tsinghua University, Beijing 100084, China}

\author{Pu Yu}
\author{Xu-Cun Ma}
\email[]{xucunma@mail.tsinghua.edu.cn}
\affiliation{State Key Laboratory of Low-Dimensional Quantum Physics, Department of Physics, Tsinghua University, Beijing 100084, China}
\affiliation{Frontier Science Center for Quantum Information, Beijing 100084, China}
\author{Qi-Kun Xue}
\email[]{qkxue@mail.tsinghua.edu.cn}
\affiliation{State Key Laboratory of Low-Dimensional Quantum Physics, Department of Physics, Tsinghua University, Beijing 100084, China}
\affiliation{Frontier Science Center for Quantum Information, Beijing 100084, China}
\affiliation{Beijing Academy of Quantum Information Sciences, Beijing 100193, China}
\affiliation{Southern University of Science and Technology, Shenzhen 518055, China}
\author{Can-Li Song}
\email[]{clsong07@mail.tsinghua.edu.cn}
\affiliation{State Key Laboratory of Low-Dimensional Quantum Physics, Department of Physics, Tsinghua University, Beijing 100084, China}
\affiliation{Frontier Science Center for Quantum Information, Beijing 100084, China}

\begin{abstract}
Electron-doped infinite-layer Sr$_{1-x}$Eu$_{x}$CuO$_{2+y}$ films over a wide doping range have been prepared epitaxially on SrTiO$_3$(001) using reactive molecular beam epitaxy. In-plane transport measurements of the single crystalline samples reveal a dome-shaped nodeless superconducting phase centered at $x$ $\sim$ 0.15, a Fermi-liquid behavior and pronounced upturn in low temperature resistivity. We show that the resistivity upturn follows square-root temperature dependence, suggesting the emergence of superconductivity via a three-dimensional percolation process. The percolative superconductivity is corroborated  spectroscopically by imaging the electronic phase separation between superconducting and metallic phases with low-temperature scanning tunneling microscopy. Furthermore, we visualize interstitial and apical oxygen anions that rapidly increase in number as $x>$ 0.12, and elucidate their impacts on the superconductivity and normal-state resistivity.
\end{abstract}

\maketitle
\begin {spacing}{1.01}
The microscopic mechanism of high-temperature superconductivity (SC) in cuprates is a long-standing mystery in condensed matter physics \cite{keimer2015quantum}. Searching for more materials with analogous traits to cuprates has appealed considerable research interest \cite{kim2014fermi, zhong2016nodeless, li2019superconductivity}. It is well known that the cuprate superconductivity emerges upon doping the quintessential CuO$_2$ planes with electrons or holes. A comparative study of the two opposite doping regimes may serve as a convenient avenue towards understanding the cuprate superconductors. However, the available data of cuprates, so far, has been highly clustered in the hole-doped side \cite{fischer2007scanning, chu2015hole}. Much fewer measurements of electron-doped cuprates are limited to the $\textrm{T}'$ family of Ln$_{2-x}$Ce$_x$CuO$_4$ (Ln = La, Nd, Pr, Sm, Eu and Gd) \cite{Armitage2002doping, Armitage2010progress, niestemski2007distinct, fournier2015t, da2015charge, greene2020strange}. No consensus has been reached on commonalities or distinctions between the hole- and electron-doped compounds \cite{weber2010strength, li2019hole}. Furthermore, the material synthesis of $\textrm{T}'$-type cuprates is complicated by unwanted competing oxide phases that may yield erroneous conclusions on their physical properties \cite{Mang2004phase}.

Infinite-layer compounds of $A$CuO$_2$ ($A$ = Ca, Sr, Ba) have the simplest crystal structure among all cuprates and become superconducting after doping electrons by partial substitution of trivalent metal cations for $A^{2+}$ \cite{smith1991electron, karimoto2001single, Zapf2005dimensionality}. Regarding to the CuO$_2$ surface termination, this family of cuprates is admirably adapted to explore the essence of the copper-oxide superconductors \cite{Harter2012nodeless, zhong2020direct, fan2021direct}. A problem is that they are metastable and pose significant challenges for single crystal synthesis. To date, only limited studies have been conducted on the epitaxial films of $A$CuO$_2$ doped with La, Nd, Sm and Gd \cite{smith1991electron, karimoto2001single, Zapf2005dimensionality, Harter2012nodeless, zhong2020direct, fan2021direct, ikeda1993synthesis}, with a limited range of electron doping due to the low solubility of these lanthanides. This holds true especially for the infinite-layer films on SrTiO$_3$(001) \cite{karimoto2001single, Harter2012nodeless, zhong2020direct, fan2021direct, Molecular2020fan}, in which the interstitial apical oxygens occupy the apical position (O$_\textrm{A}$), contribute holes and become overwhelming with increasing trivalent dopants, posing a constraint for drawing an electron doping-dependent phase diagram. Here we succeed to dope the infinite-layer SrCuO$_2$ films on the SrTiO$_3$(001) substrates with the previously unexplored Eu cations [Fig.\ 1(a)], and show that the smaller ionic radius of Eu$^{3+}$ greatly reduces the intake of O$_\textrm{A}$ in the films. In sharp contrast to Sr$_{1-x}$La$_{x}$CuO$_{2}$ and Sr$_{1-x}$Nd$_{x}$CuO$_{2}$ \cite{zhong2020direct, fan2021direct}, this allows us to eliminate the second O$_\textrm{A}$-ordered phase above $x$ $\sim$ 0.100, first prepare the electron-doped infinite-layer Sr$_{1-x}$Eu$_{x}$CuO$_{2+y}$ (SECO) single crystalline films over a wide doping range (0.062 $<x<$ 0.246) and establish a complete superconducting phase diagram. We further unravel a percolative nature of the superconductivity, and elucidate the effects of O$_\textrm{A}$ remnants on the properties of SECO by combining electrical transport measurements and \textit{in-situ} scanning tunneling microscopy (STM).

Our experiments were conducted in two ozone-assisted molecular beam epitaxy (O-MBE) systems, one of which is connected to an ultrahigh vacuum (UHV) low temperature STM (Unisoku) for \textit{in-situ} atomic-scale characterization at 4.8 K. Semi-insulating and Nb-doped (0.5\%wt) SrTiO$_3$(001) substrates were used to grow the SECO films for electrical transport and STM measurements, respectively. The MBE growth of SECO films with various Eu content $x$ is similar to that of La and Nd-doped SrCuO$_2$ \cite{zhong2020direct, Molecular2020fan}, with $x$ determined by measuring the flux ratio between the Eu and Cu sources via a quartz crystal microbalance (Inficon SQM160H). All STM topographies were acquired in constant current mode with the voltage $V$ applied to the sample. The differential conductance d$I$/d$V$ curves and maps were measured by using a standard lock-in technique with a small bias modulation at 937 Hz. After \textit{ex-situ} x-ray diffraction (XRD) measurements using the monochromatic Cu $K_{\alpha1}$ radiation with a wavelength of 1.5406 \AA, we measured the electrical resistivity via a four-terminal configuration in a commercial physical property measurement system.
\end {spacing}

\begin{figure}[t]
\includegraphics[width=\columnwidth]{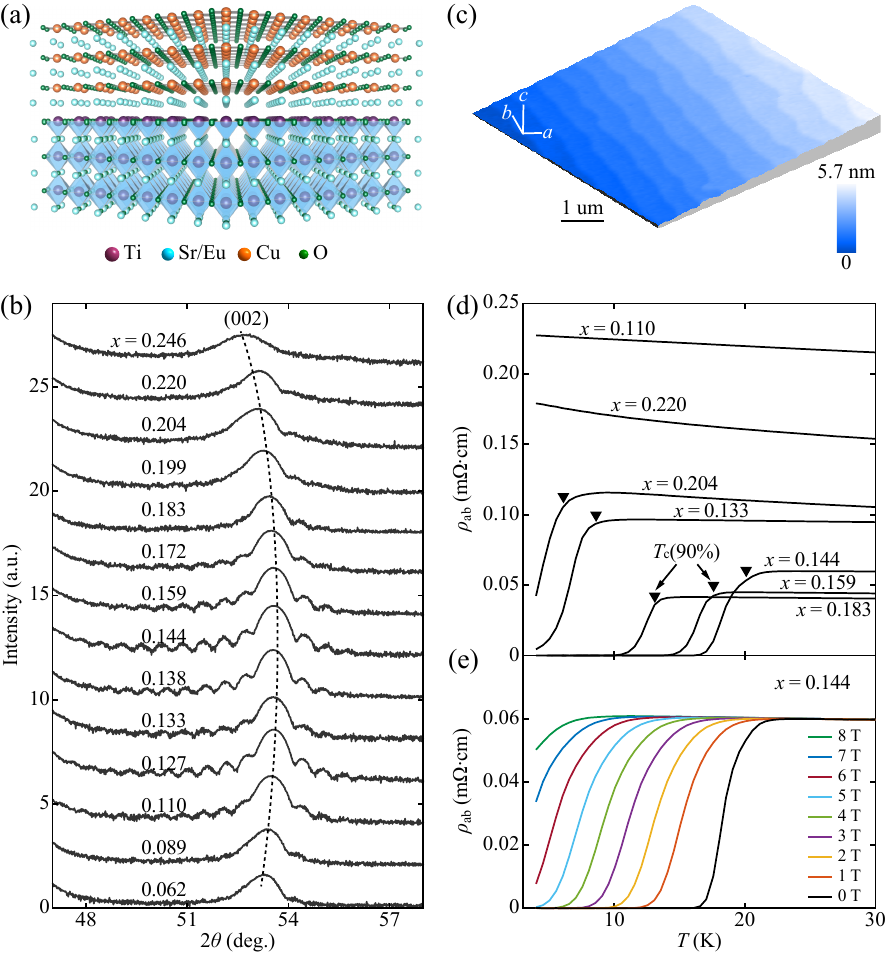}
\caption{(a) Sketch of epitaxial infinite-layer SECO thin films on SrTiO$_3$(001). (b) XRD patterns around the (002) diffraction peaks of SECO films with different Eu contents $x$, as indicated. The dashed gray curve highlights an evolution of the SECO(002) peaks. (c) AFM topography (5 um $\times$ 5 um) of the SECO films. (d) Temperature-dependent electrical resistivity for SECO films at varied $x$, with black triangles denoting the $T_{c}$ onsets at the 90\% normal-state resistivity. (e) Suppression of superconductivity by magnetic fields for $x$ = 0.144 SECO films.
}
\end{figure}

\begin{figure}[h]
\includegraphics[width=\columnwidth]{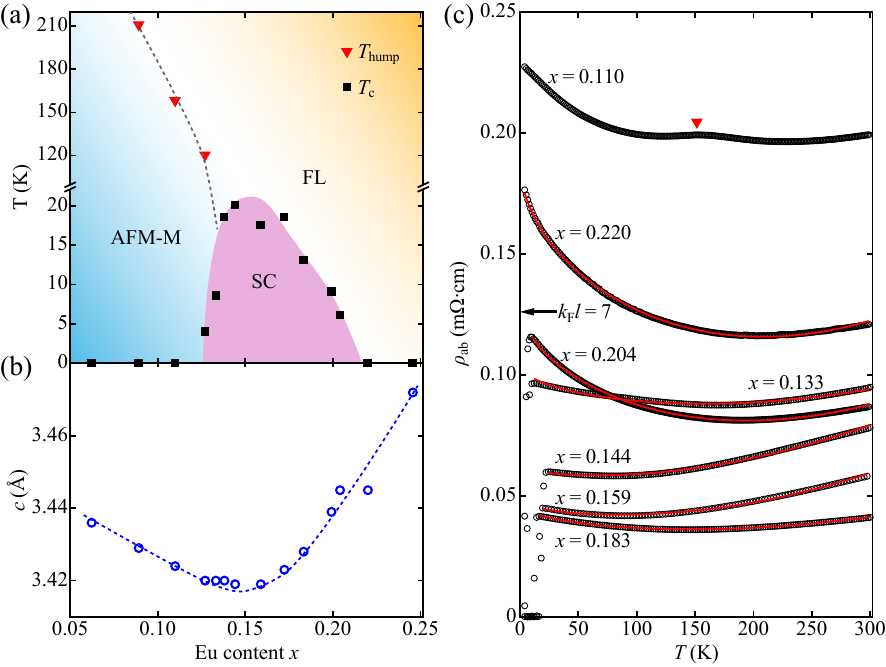}
\caption{(a) Electronic phase diagram showing the evolution of $T_\textrm{hump}$ (red triangles) and $T_\textrm{c}$ (black squares) as a function of the Eu doping content $x$. (b) Non-monotonic variation of the out-of-plane lattice constant $c$ with increasing $x$ in SECO. The dashed line is a guide to the eye. (c) Extended resistivity versus temperature plots for SECO films with various Eu doping $x$ (black circles) and fits of the normal-state resistivity to Eq.\ (1) (red curves), presenting a universal low-temperature resistivity upturn and hidden Fermi-liquid behavior. The red triangle marks $T_\textrm{hump}$ hallmark of possible coexistence of itinerant and localized electron carries.
}
\end{figure}

Figure 1 (b) depicts a series of XRD data of single crystalline SECO thin films at various Eu contents $x$, where the inevitable O$_\textrm{A}$ has been minimized by a reduction annealing in UHV. It is apparent that all SECO thin films are in tetragonal phase and there is no peak related to the interstitial O$_\textrm{A}$-rich and ordered phase until $x >$ 0.240 (also see Fig.\ S1 in the Supplemental Material) \cite{supplementary}, in stark contrast to the (La, Nd)-doped SrCuO$_2$ cases \cite{karimoto2001single, zhong2020direct, Molecular2020fan}. Kiessig fringes were commonly observed, from which we estimate the film thickness to be $\sim$ 16.5 nm. This indicates a high-quality epitaxial SECO films with uniform thickness, as evident by atomic step-terrace morphology in atomic force microscopy (AFM) image [Fig.\ 1(c)]. Figure 1(d) plots the temperature dependence of in-plane resistivity $\rho_{ab}$ for seven samples. A phase transition from insulator to superconductor is found around $x \sim$ 0.127 [Fig.\ S2]. For $x >$ 0.204, however, the SECO films revert to a metallic-like state ($d\rho_{ab}$/$dT$ $>$ 0). In between, the samples display apparent superconducting phase transition into zero-resistivity state, with a critical temperature $T_\textrm{c}$ up to 20 K. The superconducting state is gradually suppressed by $c$-axis aligned magnetic field in Fig.\ 1(e).

Importantly, the resistivities for the SECO films at $x \leq$ 0.127 exhibit atypical humps marked by red triangles in Fig.\ S2. These intriguing resistivity humps, reminiscent of the underdoped $\textrm{T}'$-type Nd$_{2-x}$Ce$_x$CuO$_4$ \cite{Onose2004charge}, pressurized alkaline iron selenide superconductors \cite{Gao2014role} and antiferromagnetic (AFM) insulator LaMnPO \cite{guo2013observation}, originate from partial electronic delocalization driven by the Eu dopants and disappear at elevated $x$. In Figs.\ 2(a) and 2(b), we summarize the extracted resistivity hump temperature $T_\textrm{hump}$, $T_\textrm{c}$ and out-of-plane lattice constant $c$ as a function of the Eu content $x$. Evidently, $T_\textrm{hump}$ declines with increasing $x$, whereas the $T_\textrm{c}$ initially increases and then decreases to zero, forming a dome-shaped superconducting phase diagram. Compared with the hole-doped cuprates, the electron-doped SECO films exhibit a relatively narrow doping range of $x\sim$ 0.13$-$0.21 for superconductivity and an optimal doping at $x$ $\sim$ 0.15 \cite{keimer2015quantum, chu2015hole, Armitage2010progress}. It is clear that the optimal doping of $x$ $\sim$ 0.15 coincides with a minimum in $c$ [Fig.\ 2(b)]. Such a coincidence and the nonmonotonic variation of $c$ with $x$ correlate with the O$_\textrm{A}$ remnants in SECO, which we will explain later.

In addition, the resistivity $\rho_{ab}$ shows a pronounced upturn at low temperatures that alters little with magnetic field [Fig.\ 1(d)], even for the superconducting SECO films [Fig.\ 1(e)]. This is more clearly seen from the resistivity curves over an extended temperature range in Fig.\ 2(c). Such an insulating upturn in resistivity has been widely identified in other cuprate superconductors and often follows a puzzling log(1/$T$) divergence behavior \cite{Ando1995logarithmic,Boebinger1996insulator, Fournier1998insulator,Ono2000metal}, although a mysterious deviation sometimes occurs at lower temperatures \cite{Ono2000metal, Sekitani2003Kondo, wang2005transport}. In SECO, we observe a similar logarithmic-like upturn in low temperature resistivity including the superconducting ones [Fig.\ S3(a)], with some samples showing a resistivity saturation behavior toward zero temperature (namely $x$ = 0.133, 0.204 and 0.220). A closer inspection shows that the low temperature $\rho_{ab}$ can be better described by a square-root temperature dependence [Fig.\ S3(b)], whereas the $\rho_{ab}$ scales quadratically with $T$ at high temperatures [Fig.\ S3(c)]. Indeed, a combination of both the square-root and quadratic terms (see the red curves in Fig.\ 2(c)) nicely fits the normal-state $\rho_{ab}$ over almost two decades of temperature via
\begin{equation}
\rho_{ab}= \rho_{0}-A_{\textrm{sq}}\sqrt{T}+A_{2}T^2,
\end{equation}
where $\rho_{0}$ is the temperature-independent resistivity, the second and third terms come from electron-electron interactions. The quadratic temperature dependence of $\rho_{ab}$ indicates a hidden Fermi liquid (FL) behavior in SECO [Fig.\ 2(a)], which seems generic for both electron- and hole-doped cuprates \cite{Li2016hidden, barivsic2019evidence, pelc2019unusual}.

The resistivity upturn and saturation behaviors in low temperatures have been tentatively discussed in terms of the Kondo effect \cite{Sekitani2003Kondo}, two-dimensional (2D) weak localization (WL) by disorder \cite{Lee1985disordered}, unusual electron-electron interactions \cite{Beloborodov2003transport} and even a $d$-wave superconducting order scenario \cite{Zhou2018logarithmic}. The WL and Kondo effect are largely suppressed under magnetic field. In SECO, however, the resistivity upturn has nothing to do with the applied field [Fig.\ S4], which rules out both WL and Kondo mechanisms for the low temperature upturn in $\rho_{ab}$. In fact, the insulator-like low-$T$ resistivity upturn ($d\rho_{ab}$/$dT$ $<$ 0) has been observed in the superconducting SECO films with extremely low $\rho_{ab}$, from which we estimate $k_\textrm{F}\ell>$ 7 in a free electron model for 2D material \cite{Boebinger1996insulator}, with $k_\textrm{F}$ and $\ell$ representing the Fermi wave vector and the mean free path between disorder-induced scattering events, respectively. Instead, the cusplike $\sqrt{T}$ behavior of resistivity has been extensively discussed in granular electronic systems and roots at 3D electron-electron interactions, in which the coherent electron motion on the scales larger than the granule size dominates the charge transport \cite{Beloborodov2007grnular, Efetov2003coulomb, Beloborodov2003transport, Sachser2011universal}. This points toward granular metallicity due to the intrinsic inhomogeneity of SECO, from which the superconductivity emerges via a 3D percolation process. Actually, the percolation picture agrees gently with the remarkable electronic inhomogeneity observed in high-$T_\textrm{c}$ superconductors \cite{pan2001microscopic, parra2021signatures} and has been recently employed to explain the unusual behaviors of cuprates \cite{pelc2019unusual, pelc2018emergence, popvcevic2018percolative, Yu2019universal}.

\begin{figure}[t]
\includegraphics[width=\columnwidth]{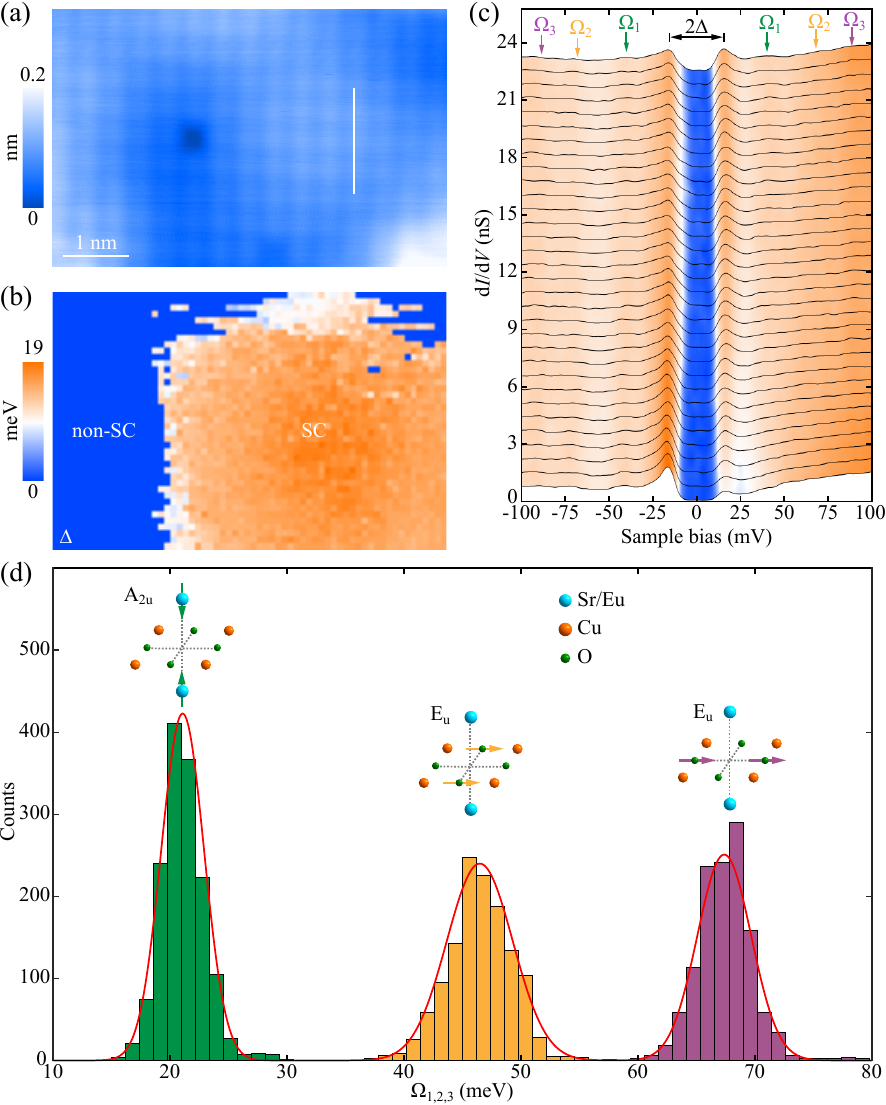}
\caption{(a) STM topography of SECO films (60 $\textrm{\AA}$ $\times$ 38 $\textrm{\AA}$, $V$ = -0.9 V, $I$= 20 pA). (b) Spatial map of  $\Delta$ (half the distance between coherence peaks) showing an electronic segregation between SC and non-SC phases. (c) Tunneling spectra taken at equal separations (0.65 $\textrm{\AA}$) along the white solid line in (a). Outside the superconducting gap, the peak-dip-hump features develop from three distinct vibrational phonons $\Omega_{\textrm{1,2,3}}$ and are marked by colored down arrows. Setpoint: \textit{V} = -100 mV and \textit{I} = 100 pA. (d) Histogram of $\Omega_{\textrm{1,2,3}}$ measured from the d$I$/d$V$ spectra in the superconducting domain in (b).
}
\end{figure}

The percolative superconductivity has been confirmed by spatially dependent tunneling spectra in various SECO films. Without loss of generality, the STM topographic images for all SECO samples are electronically separated into bright and dark domains with various electron doping [Fig.\ S5], primarily driven by an inhomogeneous distribution of Eu dopants. In domains with moderate doping levels, the superconductivity emerges as local electron pairing, accompanied by fully opened superconducting gaps at the Fermi level ($E_\textrm{F}$). One representative dataset acquired on the superconducting SECO films is exemplified in Figs.\ 3(a)-3(c). By measuring the spatial dependence of d$I$/d$V$ spectra at low energies, we found a superconducting puddle [Fig.\ 3(b)] in which the nodeless pairing gaps are robustly observed [Fig.\ 3(c)]. In non-superconducting regions [Fig.\ S6], however, the d$I$/d$V$ spectra are characteristic of metallic-like features and match with the electrical transport measurements in Fig.\ 2(c). This significantly resembles the Nd-doped SrCuO$_2$ films \cite{fan2021direct}, except that the superconducting gap turns out to be a little more homogeneously distributed in spatial and exhibits a smaller mean magnitude of $\Delta_{\textrm{mean}}$ $\sim$ 18 meV [Fig.\ S7]. This fact, in conjunction with the frequent occurrence of superconducting gaps in topographically relatively faint regions [Figs.\ 3(a) and 3(b)], helps reach the percolation threshold for zero-resistivity superconducting state in SECO [Fig.\ 1(d)]. Such local imaging of granular superconductivity constitutes convincing support for the emergence of superconductivity in SECO via a percolation process. The strong agreement between microscopic electronic features and macroscopic transport hints that the STM probe of the topmost CuO$_2$ planes largely reflects the nature of bulk superconductivity in SECO.

\begin{figure}[t]
\includegraphics[width=\columnwidth]{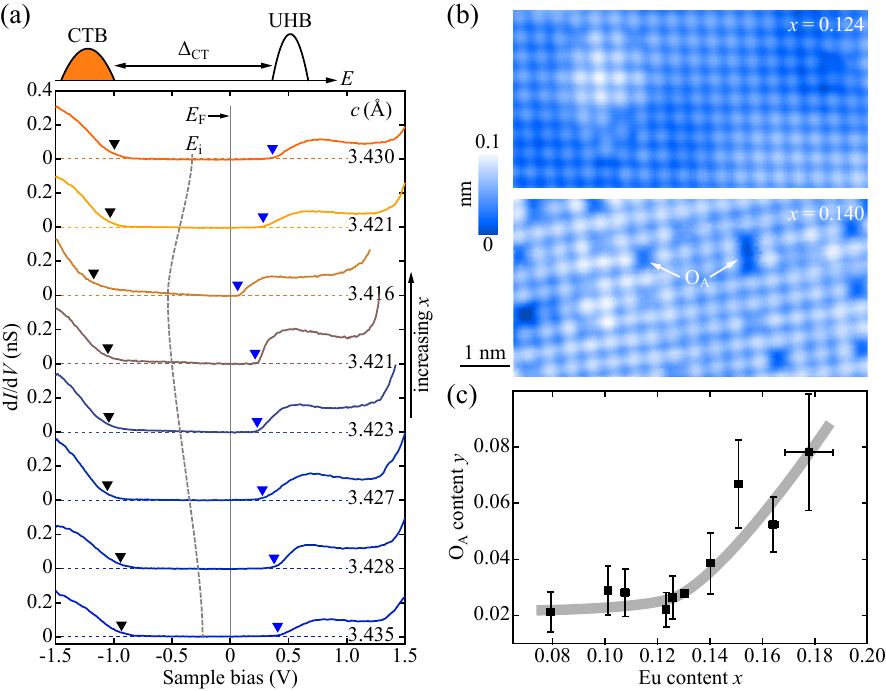}
\caption{(a) Spatially averaged d$I$/d$V$ spectra ($V$ = -1.5 V, $I$ = 100 pA) showing a non-monotonic dependence of the Mott-Hubbard band structure of SECO on the Eu content $x$. The gray solid and dashed lines denote $E_\textrm{F}$ and $E_\textrm{i}$, respectively. Top panel:\ schematic energy bands of cuprates showing only CTB (orange) and UHB (unfilled). (b) STM topographies (70 $\textrm{\AA}$ $\times$ 35 $\textrm{\AA}$, $I$ = 20 pA) of two different SECO films with extra O$_\textrm{A}$ defects appearing as dark dots. Top panel:\ $x$ = 0.124, $V$ = -1.5 V; Bottom panel:\ $x$ = 0.140, $V$ = 1.2 V. (c) Statistical measure of O$_\textrm{A}$ content $y$ as a function of $x$. The gray thick line is guide to the eye, while the error bars mark the standard deviations of $x$ and $y$ values obtained from various regions and samples.
}
\end{figure}

In analogy to Nd-doped SrCuO$_2$ \cite{fan2021direct}, we also observe multiple peak-dip-hump structures outside the superconducting gaps in SNCO, which stem from three vibrational phonons $\Omega_{\textrm{1,2,3}}$ marked by the colored down arrows in Fig.\ 3(c). By measuring their mode energies from thousands of superconducting d$I$/d$V$ spectra via a well-established method (e.g.\ Fig.\ S8) \cite{fan2021direct}, we estimate the average mode energies to be $\Omega_{\textrm{1}}$ = 21 $\pm$ 2 meV (external mode), $\Omega_{\textrm{2}}$ = 46 $\pm$ 3 meV (bending mode) and $\Omega_{\textrm{3}}$ = 68 $\pm$ 2 meV (stretching mode), respectively. The statistical errors of $\Omega_{\textrm{1,2,3}}$ indicate the full width at half maximum of the corresponding Gaussian peaks (red curve) in Fig.\ 3(d). As anticipated, the lattice vibrational energies $\Omega_{\textrm{1,2,3}}$ depend little on $\Delta$ and the local ratio of $\Omega_{\textrm{1,2,3}}$ to 2$\Delta$ ($\Omega_{\textrm{1,2,3}}$/2$\Delta$) can exceed unity [Fig.\ S9],
running counter to the spin excitation scenario for $\Omega_{\textrm{1,2,3}}$ \cite{fan2021direct}. In SECO, $\Omega_{\textrm{3}}$ = 68 meV appears to be slightly smaller than that (72 meV) in Nd-doped SrCuO$_2$ \cite{fan2021direct}. This is understandable because the stretching phonon mode $\Omega_{\textrm{3}}$ is very sensitive to the Cu-O distance in cuprates \cite{tajima1990optical, Tajima1991optical}.The heavier electron doping of SECO stretches the CuO$_2$ bonds more appreciably and causes a tiny redshift of $\Omega_{\textrm{3}}$ \cite{Armitage2010progress, fournier2015t, karimoto2001single}. These findings establish the universal phonon modes that are strongly coupled with the electrons and induce superconductivity in the infinite-layer cuprates.

To provide insight into the unusual doping dependence of the out-of-plane lattice constant $c$ in Fig.\ 2(b), we have measured a series of wide-energy-ranged tunneling spectra ($\pm$ 1.5 eV) in the SECO films at various Eu contents, as illustrated in Fig.\ 4(a). In analogy to the (La, Nd)-doped cases \cite{zhong2020direct, fan2021direct, Molecular2020fan}, the fundamental Mott-Hubbard band structure of the cuprate CuO$_2$ planes, characteristic of a charge-transfer gap between charge-transfer band (CTB, black triangles) and upper-Hubbard band (UHB, blue triangles), remains essentially unchanged against the Eu doping. However, the $E_\textrm{F}$ initially moves upwards and then downwards relative to the midgap energy $E_\textrm{i}$ with increasing $x$. Given that the $E_\textrm{F}$-$E_\textrm{i}$ proves as a good indicator of doping level \cite{zhong2020direct, fan2021direct, Molecular2020fan, Wang2020electronic}, the downward shift of $E_\textrm{F}$ indicates additional involvement of hole carries. By imaging atomic-scale defects in SECO [Fig.\ 4(b)], we find that they are the apical oxygens O$_\textrm{A}$ justly beneath the Cu atom that create the extra holes \cite{zhong2020direct}. Although the O$_\textrm{A}$-ordered phase is eliminated, a tiny amount of O$_\textrm{A}$ rapidly increase in number as $x>$ 0.12 [Fig.\ 4(c)]. As electron acceptors, these O$_\textrm{A}$ remnants compensate for the electron doping by Eu, reduce the net electron carriers in SECO, and expand the lattice, matching nicely with the increase of $c$ as $x>$ 0.15 [Figs.\ 2(b) and 4(a)].

We comment on the impacts of O$_\textrm{A}$ remnants on superconductivity and the normal-state resistivity. First, they reduce the effective electron doping by Eu and delay the emergence of superconductivity in the electron-doped SECO films at $x>$ 0.12 [Fig.\ 1(d)]. Second, due to the reduced electron carriers the normal-state $\rho_{ab}$ is sharply increased at $x>$ 0.18, followed by an abrupt frustration of the superconductivity [Fig.\ 2(c)]. The two results yield the narrow superconducting phase diagram in Fig.\ 2(a). Last but not least, the O$_\textrm{A}$ is more apparently increased above $x\sim$ 0.12 [Fig.\ 4(c)] that slightly deviates from $x \sim$ 0.15 for the optimal superconductivity. This discrepancy is understandable because the $T_\textrm{c}$ of percolative superconductor is controlled by the optimal paths with the larger $x$ and smaller $y$, rather than the mean values of them, in the underdoped regime. Under this context, a removal of the O$_\textrm{A}$ remnants may further enhance superconductivity in electron-doped infinite-layer SECO films.

In summary, we have overcome the biggest material challenge of infinite-layer cuprates by successfully preparing the epitaxial SNCO films over a wide electron doping range and rarely established a complete phase diagram by combining \textit{in-situ} STM, \textit{ex-situ} XRD and electrical measurements. Our observations of the dome-shaped nodeless superconducting phase, percolative nature of the superconductivity and hidden FL behaviors shed important light on the culprit of superconductivity in cuprates. The direct visualization of interstitial O$_\textrm{A}$ remnants points to a route to enhance superconducting critical temperature in electron-doped infinite-layer cuprates.

\begin{acknowledgments}
The work was financially supported by the Ministry of Science and Technology of China (2017YFA0304600, 2018YFA0305603) and the National Natural Science Foundation of China (No.\ 12134008, 51788104). X.\ Q.\ Yu and H.\ Yan contributed equally to this work.
\end{acknowledgments}

% Create the reference section using BibTeX:
%\bibliography{SECO}

\begin{thebibliography}{50}%
\makeatletter
\providecommand \@ifxundefined [1]{%
 \@ifx{#1\undefined}
}%
\providecommand \@ifnum [1]{%
 \ifnum #1\expandafter \@firstoftwo
 \else \expandafter \@secondoftwo
 \fi
}%
\providecommand \@ifx [1]{%
 \ifx #1\expandafter \@firstoftwo
 \else \expandafter \@secondoftwo
 \fi
}%
\providecommand \natexlab [1]{#1}%
\providecommand \enquote  [1]{``#1''}%
\providecommand \bibnamefont  [1]{#1}%
\providecommand \bibfnamefont [1]{#1}%
\providecommand \citenamefont [1]{#1}%
\providecommand \href@noop [0]{\@secondoftwo}%
\providecommand \href [0]{\begingroup \@sanitize@url \@href}%
\providecommand \@href[1]{\@@startlink{#1}\@@href}%
\providecommand \@@href[1]{\endgroup#1\@@endlink}%
\providecommand \@sanitize@url [0]{\catcode `\\12\catcode `\$12\catcode
  `\&12\catcode `\#12\catcode `\^12\catcode `\_12\catcode `\%12\relax}%
\providecommand \@@startlink[1]{}%
\providecommand \@@endlink[0]{}%
\providecommand \url  [0]{\begingroup\@sanitize@url \@url }%
\providecommand \@url [1]{\endgroup\@href {#1}{\urlprefix }}%
\providecommand \urlprefix  [0]{URL }%
\providecommand \Eprint [0]{\href }%
\providecommand \doibase [0]{http://dx.doi.org/}%
\providecommand \selectlanguage [0]{\@gobble}%
\providecommand \bibinfo  [0]{\@secondoftwo}%
\providecommand \bibfield  [0]{\@secondoftwo}%
\providecommand \translation [1]{[#1]}%
\providecommand \BibitemOpen [0]{}%
\providecommand \bibitemStop [0]{}%
\providecommand \bibitemNoStop [0]{.\EOS\space}%
\providecommand \EOS [0]{\spacefactor3000\relax}%
\providecommand \BibitemShut  [1]{\csname bibitem#1\endcsname}%
\let\auto@bib@innerbib\@empty
%</preamble>
\bibitem [{\citenamefont {Keimer}\ \emph {et~al.}(2015)\citenamefont {Keimer},
  \citenamefont {Kivelson}, \citenamefont {Norman}, \citenamefont {Uchida},\
  and\ \citenamefont {Zaanen}}]{keimer2015quantum}%
  \BibitemOpen
  \bibfield  {author} {\bibinfo {author} {\bibfnamefont {B.}~\bibnamefont
  {Keimer}}, \bibinfo {author} {\bibfnamefont {S.}~\bibnamefont {Kivelson}},
  \bibinfo {author} {\bibfnamefont {M.}~\bibnamefont {Norman}}, \bibinfo
  {author} {\bibfnamefont {S.}~\bibnamefont {Uchida}}, \ and\ \bibinfo {author}
  {\bibfnamefont {J.}~\bibnamefont {Zaanen}},\ }\href {\doibase
  10.1038/nature14165} {\bibfield  {journal} {\bibinfo  {journal} {Nature}\
  }\textbf {\bibinfo {volume} {518}},\ \bibinfo {pages} {179} (\bibinfo {year}
  {2015})}\BibitemShut {NoStop}%
\bibitem [{\citenamefont {Kim}\ \emph {et~al.}(2014)\citenamefont {Kim},
  \citenamefont {Krupin}, \citenamefont {Denlinger}, \citenamefont {Bostwick},
  \citenamefont {Rotenberg}, \citenamefont {Zhao}, \citenamefont {Mitchell},
  \citenamefont {Allen},\ and\ \citenamefont {Kim}}]{kim2014fermi}%
  \BibitemOpen
  \bibfield  {author} {\bibinfo {author} {\bibfnamefont {Y.~K.}\ \bibnamefont
  {Kim}}, \bibinfo {author} {\bibfnamefont {O.}~\bibnamefont {Krupin}},
  \bibinfo {author} {\bibfnamefont {J.~D.}\ \bibnamefont {Denlinger}}, \bibinfo
  {author} {\bibfnamefont {A.}~\bibnamefont {Bostwick}}, \bibinfo {author}
  {\bibfnamefont {E.}~\bibnamefont {Rotenberg}}, \bibinfo {author}
  {\bibfnamefont {Q.}~\bibnamefont {Zhao}}, \bibinfo {author} {\bibfnamefont
  {J.~F.}\ \bibnamefont {Mitchell}}, \bibinfo {author} {\bibfnamefont {J.~W.}\
  \bibnamefont {Allen}}, \ and\ \bibinfo {author} {\bibfnamefont {B.~J.}\
  \bibnamefont {Kim}},\ }\href {\doibase 10.1126/science.1251151} {\bibfield
  {journal} {\bibinfo  {journal} {Science}\ }\textbf {\bibinfo {volume}
  {345}},\ \bibinfo {pages} {187} (\bibinfo {year} {2014})}\BibitemShut
  {NoStop}%
\bibitem [{\citenamefont {Zhong}\ \emph {et~al.}(2016)\citenamefont {Zhong},
  \citenamefont {Wang}, \citenamefont {Han}, \citenamefont {Lv}, \citenamefont
  {Wang}, \citenamefont {Zhang}, \citenamefont {Ding}, \citenamefont {Zhang},
  \citenamefont {Wang}, \citenamefont {He}, \citenamefont {Zhong},
  \citenamefont {Schneeloch}, \citenamefont {Gu}, \citenamefont {Song},
  \citenamefont {Ma},\ and\ \citenamefont {Xue}}]{zhong2016nodeless}%
  \BibitemOpen
  \bibfield  {author} {\bibinfo {author} {\bibfnamefont {Y.}~\bibnamefont
  {Zhong}}, \bibinfo {author} {\bibfnamefont {Y.}~\bibnamefont {Wang}},
  \bibinfo {author} {\bibfnamefont {S.}~\bibnamefont {Han}}, \bibinfo {author}
  {\bibfnamefont {Y.~F.}\ \bibnamefont {Lv}}, \bibinfo {author} {\bibfnamefont
  {W.~L.}\ \bibnamefont {Wang}}, \bibinfo {author} {\bibfnamefont
  {D.}~\bibnamefont {Zhang}}, \bibinfo {author} {\bibfnamefont
  {H.}~\bibnamefont {Ding}}, \bibinfo {author} {\bibfnamefont {Y.~M.}\
  \bibnamefont {Zhang}}, \bibinfo {author} {\bibfnamefont {L.}~\bibnamefont
  {Wang}}, \bibinfo {author} {\bibfnamefont {K.}~\bibnamefont {He}}, \bibinfo
  {author} {\bibfnamefont {R.~D.}\ \bibnamefont {Zhong}}, \bibinfo {author}
  {\bibfnamefont {J.~A.}\ \bibnamefont {Schneeloch}}, \bibinfo {author}
  {\bibfnamefont {G.~D.}\ \bibnamefont {Gu}}, \bibinfo {author} {\bibfnamefont
  {C.~L.}\ \bibnamefont {Song}}, \bibinfo {author} {\bibfnamefont {X.~C.}\
  \bibnamefont {Ma}}, \ and\ \bibinfo {author} {\bibfnamefont {Q.~K.}\
  \bibnamefont {Xue}},\ }\href {\doibase 10.1007/s11434-016-1145-4} {\bibfield
  {journal} {\bibinfo  {journal} {Sci. Bull.}\ }\textbf {\bibinfo {volume}
  {61}},\ \bibinfo {pages} {1239} (\bibinfo {year} {2016})}\BibitemShut
  {NoStop}%
\bibitem [{\citenamefont {Li}\ \emph {et~al.}(2019{\natexlab{a}})\citenamefont
  {Li}, \citenamefont {Zhao}, \citenamefont {Cao}, \citenamefont {Hu},
  \citenamefont {Huang}, \citenamefont {Wang}, \citenamefont {Liu},
  \citenamefont {Zhao}, \citenamefont {Zhang}, \citenamefont {Liu},
  \citenamefont {Yu}, \citenamefont {Long}, \citenamefont {Wu}, \citenamefont
  {Lin}, \citenamefont {Chen}, \citenamefont {Li}, \citenamefont {Gong},
  \citenamefont {Guguchia}, \citenamefont {Kim}, \citenamefont {Stewart},
  \citenamefont {Uemura}, \citenamefont {Uchida},\ and\ \citenamefont
  {Jin}}]{li2019superconductivity}%
  \BibitemOpen
  \bibfield  {author} {\bibinfo {author} {\bibfnamefont {W.~M.}\ \bibnamefont
  {Li}}, \bibinfo {author} {\bibfnamefont {J.~F.}\ \bibnamefont {Zhao}},
  \bibinfo {author} {\bibfnamefont {L.~P.}\ \bibnamefont {Cao}}, \bibinfo
  {author} {\bibfnamefont {Z.}~\bibnamefont {Hu}}, \bibinfo {author}
  {\bibfnamefont {Q.~Z.}\ \bibnamefont {Huang}}, \bibinfo {author}
  {\bibfnamefont {X.~C.}\ \bibnamefont {Wang}}, \bibinfo {author}
  {\bibfnamefont {Y.}~\bibnamefont {Liu}}, \bibinfo {author} {\bibfnamefont
  {G.~Q.}\ \bibnamefont {Zhao}}, \bibinfo {author} {\bibfnamefont
  {J.}~\bibnamefont {Zhang}}, \bibinfo {author} {\bibfnamefont {Q.~Q.}\
  \bibnamefont {Liu}}, \bibinfo {author} {\bibfnamefont {R.~Z.}\ \bibnamefont
  {Yu}}, \bibinfo {author} {\bibfnamefont {Y.~W.}\ \bibnamefont {Long}},
  \bibinfo {author} {\bibfnamefont {H.}~\bibnamefont {Wu}}, \bibinfo {author}
  {\bibfnamefont {H.~J.}\ \bibnamefont {Lin}}, \bibinfo {author} {\bibfnamefont
  {C.~T.}\ \bibnamefont {Chen}}, \bibinfo {author} {\bibfnamefont
  {Z.}~\bibnamefont {Li}}, \bibinfo {author} {\bibfnamefont {Z.~Z.}\
  \bibnamefont {Gong}}, \bibinfo {author} {\bibfnamefont {Z.}~\bibnamefont
  {Guguchia}}, \bibinfo {author} {\bibfnamefont {J.~S.}\ \bibnamefont {Kim}},
  \bibinfo {author} {\bibfnamefont {G.~R.}\ \bibnamefont {Stewart}}, \bibinfo
  {author} {\bibfnamefont {Y.~J.}\ \bibnamefont {Uemura}}, \bibinfo {author}
  {\bibfnamefont {S.}~\bibnamefont {Uchida}}, \ and\ \bibinfo {author}
  {\bibfnamefont {C.~Q.}\ \bibnamefont {Jin}},\ }\href {\doibase
  10.1073/pnas.1900908116} {\bibfield  {journal} {\bibinfo  {journal} {Proc.
  Natl. Acad. Sci. USA}\ }\textbf {\bibinfo {volume} {116}},\ \bibinfo {pages}
  {12156} (\bibinfo {year} {2019}{\natexlab{a}})}\BibitemShut {NoStop}%
\bibitem [{\citenamefont {Fischer}\ \emph {et~al.}(2007)\citenamefont
  {Fischer}, \citenamefont {Kugler}, \citenamefont {Maggio-Aprile},
  \citenamefont {Berthod},\ and\ \citenamefont {Renner}}]{fischer2007scanning}%
  \BibitemOpen
  \bibfield  {author} {\bibinfo {author} {\bibfnamefont {{\O}.}~\bibnamefont
  {Fischer}}, \bibinfo {author} {\bibfnamefont {M.}~\bibnamefont {Kugler}},
  \bibinfo {author} {\bibfnamefont {I.}~\bibnamefont {Maggio-Aprile}}, \bibinfo
  {author} {\bibfnamefont {C.}~\bibnamefont {Berthod}}, \ and\ \bibinfo
  {author} {\bibfnamefont {C.}~\bibnamefont {Renner}},\ }\href {\doibase
  10.1103/RevModPhys.79.353} {\bibfield  {journal} {\bibinfo  {journal} {Rev.
  Mod. Phys.}\ }\textbf {\bibinfo {volume} {79}},\ \bibinfo {pages} {353}
  (\bibinfo {year} {2007})}\BibitemShut {NoStop}%
\bibitem [{\citenamefont {Chu}\ \emph {et~al.}(2015)\citenamefont {Chu},
  \citenamefont {Deng},\ and\ \citenamefont {Lv}}]{chu2015hole}%
  \BibitemOpen
  \bibfield  {author} {\bibinfo {author} {\bibfnamefont {C.~W.}\ \bibnamefont
  {Chu}}, \bibinfo {author} {\bibfnamefont {L.~Z.}\ \bibnamefont {Deng}}, \
  and\ \bibinfo {author} {\bibfnamefont {B.}~\bibnamefont {Lv}},\ }\href
  {\doibase 10.1016/j.physc.2015.02.047} {\bibfield  {journal} {\bibinfo
  {journal} {Physica C}\ }\textbf {\bibinfo {volume} {514}},\ \bibinfo {pages}
  {290} (\bibinfo {year} {2015})}\BibitemShut {NoStop}%
\bibitem [{\citenamefont {Armitage}\ \emph {et~al.}(2002)\citenamefont
  {Armitage}, \citenamefont {Ronning}, \citenamefont {Lu}, \citenamefont {Kim},
  \citenamefont {Damascelli}, \citenamefont {Shen}, \citenamefont {Feng},
  \citenamefont {Eisaki}, \citenamefont {Shen}, \citenamefont {Mang},
  \citenamefont {Kaneko}, \citenamefont {Greven}, \citenamefont {Onose},
  \citenamefont {Taguchi},\ and\ \citenamefont {Tokura}}]{Armitage2002doping}%
  \BibitemOpen
  \bibfield  {author} {\bibinfo {author} {\bibfnamefont {N.~P.}\ \bibnamefont
  {Armitage}}, \bibinfo {author} {\bibfnamefont {F.}~\bibnamefont {Ronning}},
  \bibinfo {author} {\bibfnamefont {D.~H.}\ \bibnamefont {Lu}}, \bibinfo
  {author} {\bibfnamefont {C.}~\bibnamefont {Kim}}, \bibinfo {author}
  {\bibfnamefont {A.}~\bibnamefont {Damascelli}}, \bibinfo {author}
  {\bibfnamefont {K.~M.}\ \bibnamefont {Shen}}, \bibinfo {author}
  {\bibfnamefont {D.~L.}\ \bibnamefont {Feng}}, \bibinfo {author}
  {\bibfnamefont {H.}~\bibnamefont {Eisaki}}, \bibinfo {author} {\bibfnamefont
  {Z.~X.}\ \bibnamefont {Shen}}, \bibinfo {author} {\bibfnamefont {P.~K.}\
  \bibnamefont {Mang}}, \bibinfo {author} {\bibfnamefont {N.}~\bibnamefont
  {Kaneko}}, \bibinfo {author} {\bibfnamefont {M.}~\bibnamefont {Greven}},
  \bibinfo {author} {\bibfnamefont {Y.}~\bibnamefont {Onose}}, \bibinfo
  {author} {\bibfnamefont {Y.}~\bibnamefont {Taguchi}}, \ and\ \bibinfo
  {author} {\bibfnamefont {Y.}~\bibnamefont {Tokura}},\ }\href {\doibase
  10.1103/PhysRevLett.88.257001} {\bibfield  {journal} {\bibinfo  {journal}
  {Phys. Rev. Lett.}\ }\textbf {\bibinfo {volume} {88}},\ \bibinfo {pages}
  {257001} (\bibinfo {year} {2002})}\BibitemShut {NoStop}%
\bibitem [{\citenamefont {Armitage}\ \emph {et~al.}(2010)\citenamefont
  {Armitage}, \citenamefont {Fournier},\ and\ \citenamefont
  {Greene}}]{Armitage2010progress}%
  \BibitemOpen
  \bibfield  {author} {\bibinfo {author} {\bibfnamefont {N.~P.}\ \bibnamefont
  {Armitage}}, \bibinfo {author} {\bibfnamefont {P.}~\bibnamefont {Fournier}},
  \ and\ \bibinfo {author} {\bibfnamefont {R.~L.}\ \bibnamefont {Greene}},\
  }\href {\doibase 10.1103/RevModPhys.82.2421} {\bibfield  {journal} {\bibinfo
  {journal} {Rev. Mod. Phys.}\ }\textbf {\bibinfo {volume} {82}},\ \bibinfo
  {pages} {2421} (\bibinfo {year} {2010})}\BibitemShut {NoStop}%
\bibitem [{\citenamefont {Niestemski}\ \emph {et~al.}(2007)\citenamefont
  {Niestemski}, \citenamefont {Kunwar}, \citenamefont {Zhou}, \citenamefont
  {Li}, \citenamefont {Ding}, \citenamefont {Wang}, \citenamefont {Dai},\ and\
  \citenamefont {Madhavan}}]{niestemski2007distinct}%
  \BibitemOpen
  \bibfield  {author} {\bibinfo {author} {\bibfnamefont {F.}~\bibnamefont
  {Niestemski}}, \bibinfo {author} {\bibfnamefont {S.}~\bibnamefont {Kunwar}},
  \bibinfo {author} {\bibfnamefont {S.}~\bibnamefont {Zhou}}, \bibinfo {author}
  {\bibfnamefont {S.}~\bibnamefont {Li}}, \bibinfo {author} {\bibfnamefont
  {H.}~\bibnamefont {Ding}}, \bibinfo {author} {\bibfnamefont {Z.}~\bibnamefont
  {Wang}}, \bibinfo {author} {\bibfnamefont {P.}~\bibnamefont {Dai}}, \ and\
  \bibinfo {author} {\bibfnamefont {V.}~\bibnamefont {Madhavan}},\ }\href
  {\doibase 10.1038/nature06430} {\bibfield  {journal} {\bibinfo  {journal}
  {Nature}\ }\textbf {\bibinfo {volume} {450}},\ \bibinfo {pages} {1058}
  (\bibinfo {year} {2007})}\BibitemShut {NoStop}%
\bibitem [{\citenamefont {Fournier}(2015)}]{fournier2015t}%
  \BibitemOpen
  \bibfield  {author} {\bibinfo {author} {\bibfnamefont {P.}~\bibnamefont
  {Fournier}},\ }\href {\doibase 10.1016/j.physc.2015.02.036} {\bibfield
  {journal} {\bibinfo  {journal} {Physica C}\ }\textbf {\bibinfo {volume}
  {514}},\ \bibinfo {pages} {314} (\bibinfo {year} {2015})}\BibitemShut
  {NoStop}%
\bibitem [{\citenamefont {da~Silva~Neto}\ \emph {et~al.}(2015)\citenamefont
  {da~Silva~Neto}, \citenamefont {Comin}, \citenamefont {He}, \citenamefont
  {Sutarto}, \citenamefont {Jiang}, \citenamefont {Greene}, \citenamefont
  {Sawatzky},\ and\ \citenamefont {Damascelli}}]{da2015charge}%
  \BibitemOpen
  \bibfield  {author} {\bibinfo {author} {\bibfnamefont {E.~H.}\ \bibnamefont
  {da~Silva~Neto}}, \bibinfo {author} {\bibfnamefont {R.}~\bibnamefont
  {Comin}}, \bibinfo {author} {\bibfnamefont {F.}~\bibnamefont {He}}, \bibinfo
  {author} {\bibfnamefont {R.}~\bibnamefont {Sutarto}}, \bibinfo {author}
  {\bibfnamefont {Y.}~\bibnamefont {Jiang}}, \bibinfo {author} {\bibfnamefont
  {R.~L.}\ \bibnamefont {Greene}}, \bibinfo {author} {\bibfnamefont {G.~A.}\
  \bibnamefont {Sawatzky}}, \ and\ \bibinfo {author} {\bibfnamefont
  {A.}~\bibnamefont {Damascelli}},\ }\href {\doibase 10.1126/science.1256441}
  {\bibfield  {journal} {\bibinfo  {journal} {Science}\ }\textbf {\bibinfo
  {volume} {347}},\ \bibinfo {pages} {282} (\bibinfo {year}
  {2015})}\BibitemShut {NoStop}%
\bibitem [{\citenamefont {Greene}\ \emph {et~al.}(2020)\citenamefont {Greene},
  \citenamefont {Mandal}, \citenamefont {Poniatowski},\ and\ \citenamefont
  {Sarkar}}]{greene2020strange}%
  \BibitemOpen
  \bibfield  {author} {\bibinfo {author} {\bibfnamefont {R.~L.}\ \bibnamefont
  {Greene}}, \bibinfo {author} {\bibfnamefont {P.~R.}\ \bibnamefont {Mandal}},
  \bibinfo {author} {\bibfnamefont {N.~R.}\ \bibnamefont {Poniatowski}}, \ and\
  \bibinfo {author} {\bibfnamefont {T.}~\bibnamefont {Sarkar}},\ }\href
  {\doibase 10.1146/annurev-conmatphys-031119-050558} {\bibfield  {journal}
  {\bibinfo  {journal} {Annu. Rev. Condens. Matter Phys.}\ }\textbf {\bibinfo
  {volume} {11}},\ \bibinfo {pages} {213} (\bibinfo {year} {2020})}\BibitemShut
  {NoStop}%
\bibitem [{\citenamefont {Weber}\ \emph {et~al.}(2010)\citenamefont {Weber},
  \citenamefont {Haule},\ and\ \citenamefont {Kotliar}}]{weber2010strength}%
  \BibitemOpen
  \bibfield  {author} {\bibinfo {author} {\bibfnamefont {C.}~\bibnamefont
  {Weber}}, \bibinfo {author} {\bibfnamefont {K.}~\bibnamefont {Haule}}, \ and\
  \bibinfo {author} {\bibfnamefont {G.}~\bibnamefont {Kotliar}},\ }\href
  {\doibase 10.1038/nphys1706} {\bibfield  {journal} {\bibinfo  {journal} {Nat.
  Phys.}\ }\textbf {\bibinfo {volume} {6}},\ \bibinfo {pages} {574} (\bibinfo
  {year} {2010})}\BibitemShut {NoStop}%
\bibitem [{\citenamefont {Li}\ \emph {et~al.}(2019{\natexlab{b}})\citenamefont
  {Li}, \citenamefont {Tabis}, \citenamefont {Tang}, \citenamefont {Yu},
  \citenamefont {Jaroszynski}, \citenamefont {Bari{\v{s}}i{\'c}},\ and\
  \citenamefont {Greven}}]{li2019hole}%
  \BibitemOpen
  \bibfield  {author} {\bibinfo {author} {\bibfnamefont {Y.}~\bibnamefont
  {Li}}, \bibinfo {author} {\bibfnamefont {W.}~\bibnamefont {Tabis}}, \bibinfo
  {author} {\bibfnamefont {Y.}~\bibnamefont {Tang}}, \bibinfo {author}
  {\bibfnamefont {G.}~\bibnamefont {Yu}}, \bibinfo {author} {\bibfnamefont
  {J.}~\bibnamefont {Jaroszynski}}, \bibinfo {author} {\bibfnamefont
  {N.}~\bibnamefont {Bari{\v{s}}i{\'c}}}, \ and\ \bibinfo {author}
  {\bibfnamefont {M.}~\bibnamefont {Greven}},\ }\href {\doibase
  10.1126/sciadv.aap7349} {\bibfield  {journal} {\bibinfo  {journal} {Sci.
  Adv.}\ }\textbf {\bibinfo {volume} {5}},\ \bibinfo {pages} {eaap7349}
  (\bibinfo {year} {2019}{\natexlab{b}})}\BibitemShut {NoStop}%
\bibitem [{\citenamefont {Mang}\ \emph {et~al.}(2004)\citenamefont {Mang},
  \citenamefont {Larochelle}, \citenamefont {Mehta}, \citenamefont {Vajk},
  \citenamefont {Erickson}, \citenamefont {Lu}, \citenamefont {Buyers},
  \citenamefont {Marshall}, \citenamefont {Prokes},\ and\ \citenamefont
  {Greven}}]{Mang2004phase}%
  \BibitemOpen
  \bibfield  {author} {\bibinfo {author} {\bibfnamefont {P.~K.}\ \bibnamefont
  {Mang}}, \bibinfo {author} {\bibfnamefont {S.}~\bibnamefont {Larochelle}},
  \bibinfo {author} {\bibfnamefont {A.}~\bibnamefont {Mehta}}, \bibinfo
  {author} {\bibfnamefont {O.~P.}\ \bibnamefont {Vajk}}, \bibinfo {author}
  {\bibfnamefont {A.~S.}\ \bibnamefont {Erickson}}, \bibinfo {author}
  {\bibfnamefont {L.}~\bibnamefont {Lu}}, \bibinfo {author} {\bibfnamefont
  {W.~J.~L.}\ \bibnamefont {Buyers}}, \bibinfo {author} {\bibfnamefont {A.~F.}\
  \bibnamefont {Marshall}}, \bibinfo {author} {\bibfnamefont {K.}~\bibnamefont
  {Prokes}}, \ and\ \bibinfo {author} {\bibfnamefont {M.}~\bibnamefont
  {Greven}},\ }\href {\doibase 10.1103/PhysRevB.70.094507} {\bibfield
  {journal} {\bibinfo  {journal} {Phys. Rev. B}\ }\textbf {\bibinfo {volume}
  {70}},\ \bibinfo {pages} {094507} (\bibinfo {year} {2004})}\BibitemShut
  {NoStop}%
\bibitem [{\citenamefont {Smith}\ \emph {et~al.}(1991)\citenamefont {Smith},
  \citenamefont {Manthiram}, \citenamefont {Zhou}, \citenamefont {Goodenough},\
  and\ \citenamefont {Markert}}]{smith1991electron}%
  \BibitemOpen
  \bibfield  {author} {\bibinfo {author} {\bibfnamefont {M.~G.}\ \bibnamefont
  {Smith}}, \bibinfo {author} {\bibfnamefont {A.}~\bibnamefont {Manthiram}},
  \bibinfo {author} {\bibfnamefont {J.}~\bibnamefont {Zhou}}, \bibinfo {author}
  {\bibfnamefont {J.~B.}\ \bibnamefont {Goodenough}}, \ and\ \bibinfo {author}
  {\bibfnamefont {J.~T.}\ \bibnamefont {Markert}},\ }\href {\doibase
  10.1038/351549a0} {\bibfield  {journal} {\bibinfo  {journal} {Nature}\
  }\textbf {\bibinfo {volume} {351}},\ \bibinfo {pages} {549} (\bibinfo {year}
  {1991})}\BibitemShut {NoStop}%
\bibitem [{\citenamefont {Karimoto}\ \emph {et~al.}(2001)\citenamefont
  {Karimoto}, \citenamefont {Ueda}, \citenamefont {Naito},\ and\ \citenamefont
  {Imai}}]{karimoto2001single}%
  \BibitemOpen
  \bibfield  {author} {\bibinfo {author} {\bibfnamefont {S.~I.}\ \bibnamefont
  {Karimoto}}, \bibinfo {author} {\bibfnamefont {K.}~\bibnamefont {Ueda}},
  \bibinfo {author} {\bibfnamefont {M.}~\bibnamefont {Naito}}, \ and\ \bibinfo
  {author} {\bibfnamefont {T.}~\bibnamefont {Imai}},\ }\href {\doibase
  10.1063/1.1410872} {\bibfield  {journal} {\bibinfo  {journal} {Appl. Phys.
  Lett.}\ }\textbf {\bibinfo {volume} {79}},\ \bibinfo {pages} {2767} (\bibinfo
  {year} {2001})}\BibitemShut {NoStop}%
\bibitem [{\citenamefont {Zapf}\ \emph {et~al.}(2005)\citenamefont {Zapf},
  \citenamefont {Yeh}, \citenamefont {Beyer}, \citenamefont {Hughes},
  \citenamefont {Mielke}, \citenamefont {Harrison}, \citenamefont {Park},
  \citenamefont {Kim},\ and\ \citenamefont {Lee}}]{Zapf2005dimensionality}%
  \BibitemOpen
  \bibfield  {author} {\bibinfo {author} {\bibfnamefont {V.~S.}\ \bibnamefont
  {Zapf}}, \bibinfo {author} {\bibfnamefont {N.~C.}\ \bibnamefont {Yeh}},
  \bibinfo {author} {\bibfnamefont {A.~D.}\ \bibnamefont {Beyer}}, \bibinfo
  {author} {\bibfnamefont {C.~R.}\ \bibnamefont {Hughes}}, \bibinfo {author}
  {\bibfnamefont {C.~H.}\ \bibnamefont {Mielke}}, \bibinfo {author}
  {\bibfnamefont {N.}~\bibnamefont {Harrison}}, \bibinfo {author}
  {\bibfnamefont {M.~S.}\ \bibnamefont {Park}}, \bibinfo {author}
  {\bibfnamefont {K.~H.}\ \bibnamefont {Kim}}, \ and\ \bibinfo {author}
  {\bibfnamefont {S.~I.}\ \bibnamefont {Lee}},\ }\href {\doibase
  10.1103/PhysRevB.71.134526} {\bibfield  {journal} {\bibinfo  {journal} {Phys.
  Rev. B}\ }\textbf {\bibinfo {volume} {71}},\ \bibinfo {pages} {134526}
  (\bibinfo {year} {2005})}\BibitemShut {NoStop}%
\bibitem [{\citenamefont {Harter}\ \emph {et~al.}(2012)\citenamefont {Harter},
  \citenamefont {Maritato}, \citenamefont {Shai}, \citenamefont {Monkman},
  \citenamefont {Nie}, \citenamefont {Schlom},\ and\ \citenamefont
  {Shen}}]{Harter2012nodeless}%
  \BibitemOpen
  \bibfield  {author} {\bibinfo {author} {\bibfnamefont {J.~W.}\ \bibnamefont
  {Harter}}, \bibinfo {author} {\bibfnamefont {L.}~\bibnamefont {Maritato}},
  \bibinfo {author} {\bibfnamefont {D.~E.}\ \bibnamefont {Shai}}, \bibinfo
  {author} {\bibfnamefont {E.~J.}\ \bibnamefont {Monkman}}, \bibinfo {author}
  {\bibfnamefont {Y.}~\bibnamefont {Nie}}, \bibinfo {author} {\bibfnamefont
  {D.~G.}\ \bibnamefont {Schlom}}, \ and\ \bibinfo {author} {\bibfnamefont
  {K.~M.}\ \bibnamefont {Shen}},\ }\href {\doibase
  10.1103/PhysRevLett.109.267001} {\bibfield  {journal} {\bibinfo  {journal}
  {Phys. Rev. Lett.}\ }\textbf {\bibinfo {volume} {109}},\ \bibinfo {pages}
  {267001} (\bibinfo {year} {2012})}\BibitemShut {NoStop}%
\bibitem [{\citenamefont {Zhong}\ \emph {et~al.}(2020)\citenamefont {Zhong},
  \citenamefont {Fan}, \citenamefont {Wang}, \citenamefont {Wang},
  \citenamefont {Zhang}, \citenamefont {Zhu}, \citenamefont {Dou},
  \citenamefont {Yu}, \citenamefont {Wang}, \citenamefont {Zhang},
  \citenamefont {Zhu}, \citenamefont {Song}, \citenamefont {Ma},\ and\
  \citenamefont {Xue}}]{zhong2020direct}%
  \BibitemOpen
  \bibfield  {author} {\bibinfo {author} {\bibfnamefont {Y.}~\bibnamefont
  {Zhong}}, \bibinfo {author} {\bibfnamefont {J.~Q.}\ \bibnamefont {Fan}},
  \bibinfo {author} {\bibfnamefont {R.~F.}\ \bibnamefont {Wang}}, \bibinfo
  {author} {\bibfnamefont {S.}~\bibnamefont {Wang}}, \bibinfo {author}
  {\bibfnamefont {X.~F.}\ \bibnamefont {Zhang}}, \bibinfo {author}
  {\bibfnamefont {Y.}~\bibnamefont {Zhu}}, \bibinfo {author} {\bibfnamefont
  {Z.}~\bibnamefont {Dou}}, \bibinfo {author} {\bibfnamefont {X.~Q.}\
  \bibnamefont {Yu}}, \bibinfo {author} {\bibfnamefont {Y.}~\bibnamefont
  {Wang}}, \bibinfo {author} {\bibfnamefont {D.}~\bibnamefont {Zhang}},
  \bibinfo {author} {\bibfnamefont {J.}~\bibnamefont {Zhu}}, \bibinfo {author}
  {\bibfnamefont {C.~L.}\ \bibnamefont {Song}}, \bibinfo {author}
  {\bibfnamefont {X.~C.}\ \bibnamefont {Ma}}, \ and\ \bibinfo {author}
  {\bibfnamefont {Q.~K.}\ \bibnamefont {Xue}},\ }\href {\doibase
  10.1103/PhysRevLett.125.077002} {\bibfield  {journal} {\bibinfo  {journal}
  {Phys. Rev. Lett.}\ }\textbf {\bibinfo {volume} {125}},\ \bibinfo {pages}
  {077002} (\bibinfo {year} {2020})}\BibitemShut {NoStop}%
\bibitem [{\citenamefont {Fan}\ \emph {et~al.}(2022)\citenamefont {Fan},
  \citenamefont {Yu}, \citenamefont {Cheng}, \citenamefont {Wang},
  \citenamefont {Wang}, \citenamefont {Ma}, \citenamefont {Hu}, \citenamefont
  {Zhang}, \citenamefont {Ma}, \citenamefont {Xue},\ and\ \citenamefont
  {Song}}]{fan2021direct}%
  \BibitemOpen
  \bibfield  {author} {\bibinfo {author} {\bibfnamefont {J.~Q.}\ \bibnamefont
  {Fan}}, \bibinfo {author} {\bibfnamefont {X.~Q.}\ \bibnamefont {Yu}},
  \bibinfo {author} {\bibfnamefont {F.~J.}\ \bibnamefont {Cheng}}, \bibinfo
  {author} {\bibfnamefont {H.}~\bibnamefont {Wang}}, \bibinfo {author}
  {\bibfnamefont {R.~F.}\ \bibnamefont {Wang}}, \bibinfo {author}
  {\bibfnamefont {X.~B.}\ \bibnamefont {Ma}}, \bibinfo {author} {\bibfnamefont
  {X.~P.}\ \bibnamefont {Hu}}, \bibinfo {author} {\bibfnamefont
  {D.}~\bibnamefont {Zhang}}, \bibinfo {author} {\bibfnamefont {X.~C.}\
  \bibnamefont {Ma}}, \bibinfo {author} {\bibfnamefont {Q.~K.}\ \bibnamefont
  {Xue}}, \ and\ \bibinfo {author} {\bibfnamefont {C.~L.}\ \bibnamefont
  {Song}},\ }\href {\doibase 10.1093/nsr/nwab225} {\bibfield  {journal}
  {\bibinfo  {journal} {Natl. Sci. Rev.}\ }\textbf {\bibinfo {volume} {9}},\
  \bibinfo {pages} {nwab225} (\bibinfo {year} {2022})}\BibitemShut {NoStop}%
\bibitem [{\citenamefont {Ikeda}\ \emph {et~al.}(1993)\citenamefont {Ikeda},
  \citenamefont {Hiroi}, \citenamefont {Azuma}, \citenamefont {Takano},
  \citenamefont {Bando},\ and\ \citenamefont {Takeda}}]{ikeda1993synthesis}%
  \BibitemOpen
  \bibfield  {author} {\bibinfo {author} {\bibfnamefont {N.}~\bibnamefont
  {Ikeda}}, \bibinfo {author} {\bibfnamefont {Z.}~\bibnamefont {Hiroi}},
  \bibinfo {author} {\bibfnamefont {M.}~\bibnamefont {Azuma}}, \bibinfo
  {author} {\bibfnamefont {M.}~\bibnamefont {Takano}}, \bibinfo {author}
  {\bibfnamefont {Y.}~\bibnamefont {Bando}}, \ and\ \bibinfo {author}
  {\bibfnamefont {Y.}~\bibnamefont {Takeda}},\ }\href {\doibase
  10.1016/0921-4534(93)90979-z} {\bibfield  {journal} {\bibinfo  {journal}
  {Physica C}\ }\textbf {\bibinfo {volume} {210}},\ \bibinfo {pages} {367}
  (\bibinfo {year} {1993})}\BibitemShut {NoStop}%
\bibitem [{\citenamefont {Fan}\ \emph {et~al.}(2020)\citenamefont {Fan},
  \citenamefont {Wang}, \citenamefont {Yu}, \citenamefont {Wang}, \citenamefont
  {Xiong}, \citenamefont {Song}, \citenamefont {Ma},\ and\ \citenamefont
  {Xue}}]{Molecular2020fan}%
  \BibitemOpen
  \bibfield  {author} {\bibinfo {author} {\bibfnamefont {J.~Q.}\ \bibnamefont
  {Fan}}, \bibinfo {author} {\bibfnamefont {S.~Z.}\ \bibnamefont {Wang}},
  \bibinfo {author} {\bibfnamefont {X.~Q.}\ \bibnamefont {Yu}}, \bibinfo
  {author} {\bibfnamefont {R.~F.}\ \bibnamefont {Wang}}, \bibinfo {author}
  {\bibfnamefont {Y.~L.}\ \bibnamefont {Xiong}}, \bibinfo {author}
  {\bibfnamefont {C.~L.}\ \bibnamefont {Song}}, \bibinfo {author}
  {\bibfnamefont {X.~C.}\ \bibnamefont {Ma}}, \ and\ \bibinfo {author}
  {\bibfnamefont {Q.~K.}\ \bibnamefont {Xue}},\ }\href {\doibase
  10.1103/PhysRevB.101.180508} {\bibfield  {journal} {\bibinfo  {journal}
  {Phys. Rev. B}\ }\textbf {\bibinfo {volume} {101}},\ \bibinfo {pages}
  {180508} (\bibinfo {year} {2020})}\BibitemShut {NoStop}%
\bibitem [{sup()}]{supplementary}%
  \BibitemOpen
  \href@noop {} {\bibinfo  {journal} {Details on XRD, transport and
  spatial-dependent STM measurements}\ }\BibitemShut {NoStop}%
\bibitem [{\citenamefont {Onose}\ \emph {et~al.}(2004)\citenamefont {Onose},
  \citenamefont {Taguchi}, \citenamefont {Ishizaka},\ and\ \citenamefont
  {Tokura}}]{Onose2004charge}%
  \BibitemOpen
\bibfield  {journal} {  }\bibfield  {author} {\bibinfo {author} {\bibfnamefont
  {Y.}~\bibnamefont {Onose}}, \bibinfo {author} {\bibfnamefont
  {Y.}~\bibnamefont {Taguchi}}, \bibinfo {author} {\bibfnamefont
  {K.}~\bibnamefont {Ishizaka}}, \ and\ \bibinfo {author} {\bibfnamefont
  {Y.}~\bibnamefont {Tokura}},\ }\href {\doibase 10.1103/PhysRevB.69.024504}
  {\bibfield  {journal} {\bibinfo  {journal} {Phys. Rev. B}\ }\textbf {\bibinfo
  {volume} {69}},\ \bibinfo {pages} {024504} (\bibinfo {year}
  {2004})}\BibitemShut {NoStop}%
\bibitem [{\citenamefont {Gao}\ \emph {et~al.}(2014)\citenamefont {Gao},
  \citenamefont {Yu}, \citenamefont {Sun}, \citenamefont {Wang}, \citenamefont
  {Wang}, \citenamefont {Wu}, \citenamefont {Fang}, \citenamefont {Chen},
  \citenamefont {Guo}, \citenamefont {Zhang}, \citenamefont {Gu}, \citenamefont
  {Tian}, \citenamefont {Li}, \citenamefont {Liu}, \citenamefont {Li},
  \citenamefont {Li}, \citenamefont {Jiang}, \citenamefont {Yang},
  \citenamefont {Li}, \citenamefont {Si},\ and\ \citenamefont
  {Zhao}}]{Gao2014role}%
  \BibitemOpen
  \bibfield  {author} {\bibinfo {author} {\bibfnamefont {P.}~\bibnamefont
  {Gao}}, \bibinfo {author} {\bibfnamefont {R.}~\bibnamefont {Yu}}, \bibinfo
  {author} {\bibfnamefont {L.~L.}\ \bibnamefont {Sun}}, \bibinfo {author}
  {\bibfnamefont {H.~D.}\ \bibnamefont {Wang}}, \bibinfo {author}
  {\bibfnamefont {Z.}~\bibnamefont {Wang}}, \bibinfo {author} {\bibfnamefont
  {Q.}~\bibnamefont {Wu}}, \bibinfo {author} {\bibfnamefont {M.~H.}\
  \bibnamefont {Fang}}, \bibinfo {author} {\bibfnamefont {G.~F.}\ \bibnamefont
  {Chen}}, \bibinfo {author} {\bibfnamefont {J.}~\bibnamefont {Guo}}, \bibinfo
  {author} {\bibfnamefont {C.}~\bibnamefont {Zhang}}, \bibinfo {author}
  {\bibfnamefont {D.~C.}\ \bibnamefont {Gu}}, \bibinfo {author} {\bibfnamefont
  {H.~F.}\ \bibnamefont {Tian}}, \bibinfo {author} {\bibfnamefont {J.~Q.}\
  \bibnamefont {Li}}, \bibinfo {author} {\bibfnamefont {J.}~\bibnamefont
  {Liu}}, \bibinfo {author} {\bibfnamefont {Y.~C.}\ \bibnamefont {Li}},
  \bibinfo {author} {\bibfnamefont {X.~D.}\ \bibnamefont {Li}}, \bibinfo
  {author} {\bibfnamefont {S.}~\bibnamefont {Jiang}}, \bibinfo {author}
  {\bibfnamefont {K.}~\bibnamefont {Yang}}, \bibinfo {author} {\bibfnamefont
  {A.~G.}\ \bibnamefont {Li}}, \bibinfo {author} {\bibfnamefont {Q.~M.}\
  \bibnamefont {Si}}, \ and\ \bibinfo {author} {\bibfnamefont {Z.~X.}\
  \bibnamefont {Zhao}},\ }\href {\doibase 10.1103/PhysRevB.89.094514}
  {\bibfield  {journal} {\bibinfo  {journal} {Phys. Rev. B}\ }\textbf {\bibinfo
  {volume} {89}},\ \bibinfo {pages} {094514} (\bibinfo {year}
  {2014})}\BibitemShut {NoStop}%
\bibitem [{\citenamefont {Guo}\ \emph {et~al.}(2013)\citenamefont {Guo},
  \citenamefont {Simonson}, \citenamefont {Sun}, \citenamefont {Wu},
  \citenamefont {Gao}, \citenamefont {Zhang}, \citenamefont {Gu}, \citenamefont
  {Kotliar}, \citenamefont {Aronson},\ and\ \citenamefont
  {Zhao}}]{guo2013observation}%
  \BibitemOpen
  \bibfield  {author} {\bibinfo {author} {\bibfnamefont {J.}~\bibnamefont
  {Guo}}, \bibinfo {author} {\bibfnamefont {J.~W.}\ \bibnamefont {Simonson}},
  \bibinfo {author} {\bibfnamefont {L.~L.}\ \bibnamefont {Sun}}, \bibinfo
  {author} {\bibfnamefont {Q.}~\bibnamefont {Wu}}, \bibinfo {author}
  {\bibfnamefont {P.~W.}\ \bibnamefont {Gao}}, \bibinfo {author} {\bibfnamefont
  {C.}~\bibnamefont {Zhang}}, \bibinfo {author} {\bibfnamefont {D.~C.}\
  \bibnamefont {Gu}}, \bibinfo {author} {\bibfnamefont {G.}~\bibnamefont
  {Kotliar}}, \bibinfo {author} {\bibfnamefont {M.~G.}\ \bibnamefont
  {Aronson}}, \ and\ \bibinfo {author} {\bibfnamefont {Z.~X.}\ \bibnamefont
  {Zhao}},\ }\href@noop {} {\bibfield  {journal} {\bibinfo  {journal} {Sci.
  Rep.}\ }\textbf {\bibinfo {volume} {3}},\ \bibinfo {pages} {2555} (\bibinfo
  {year} {2013})}\BibitemShut {NoStop}%
\bibitem [{\citenamefont {Ando}\ \emph {et~al.}(1995)\citenamefont {Ando},
  \citenamefont {Boebinger}, \citenamefont {Passner}, \citenamefont {Kimura},\
  and\ \citenamefont {Kishio}}]{Ando1995logarithmic}%
  \BibitemOpen
  \bibfield  {author} {\bibinfo {author} {\bibfnamefont {Y.}~\bibnamefont
  {Ando}}, \bibinfo {author} {\bibfnamefont {G.~S.}\ \bibnamefont {Boebinger}},
  \bibinfo {author} {\bibfnamefont {A.}~\bibnamefont {Passner}}, \bibinfo
  {author} {\bibfnamefont {T.}~\bibnamefont {Kimura}}, \ and\ \bibinfo {author}
  {\bibfnamefont {K.}~\bibnamefont {Kishio}},\ }\href {\doibase
  10.1103/PhysRevLett.75.4662} {\bibfield  {journal} {\bibinfo  {journal}
  {Phys. Rev. Lett.}\ }\textbf {\bibinfo {volume} {75}},\ \bibinfo {pages}
  {4662} (\bibinfo {year} {1995})}\BibitemShut {NoStop}%
\bibitem [{\citenamefont {Boebinger}\ \emph {et~al.}(1996)\citenamefont
  {Boebinger}, \citenamefont {Ando}, \citenamefont {Passner}, \citenamefont
  {Kimura}, \citenamefont {Okuya}, \citenamefont {Shimoyama}, \citenamefont
  {Kishio}, \citenamefont {Tamasaku}, \citenamefont {Ichikawa},\ and\
  \citenamefont {Uchida}}]{Boebinger1996insulator}%
  \BibitemOpen
  \bibfield  {author} {\bibinfo {author} {\bibfnamefont {G.~S.}\ \bibnamefont
  {Boebinger}}, \bibinfo {author} {\bibfnamefont {Y.}~\bibnamefont {Ando}},
  \bibinfo {author} {\bibfnamefont {A.}~\bibnamefont {Passner}}, \bibinfo
  {author} {\bibfnamefont {T.}~\bibnamefont {Kimura}}, \bibinfo {author}
  {\bibfnamefont {M.}~\bibnamefont {Okuya}}, \bibinfo {author} {\bibfnamefont
  {J.}~\bibnamefont {Shimoyama}}, \bibinfo {author} {\bibfnamefont
  {K.}~\bibnamefont {Kishio}}, \bibinfo {author} {\bibfnamefont
  {K.}~\bibnamefont {Tamasaku}}, \bibinfo {author} {\bibfnamefont
  {N.}~\bibnamefont {Ichikawa}}, \ and\ \bibinfo {author} {\bibfnamefont
  {S.}~\bibnamefont {Uchida}},\ }\href {\doibase 10.1103/PhysRevLett.77.5417}
  {\bibfield  {journal} {\bibinfo  {journal} {Phys. Rev. Lett.}\ }\textbf
  {\bibinfo {volume} {77}},\ \bibinfo {pages} {5417} (\bibinfo {year}
  {1996})}\BibitemShut {NoStop}%
\bibitem [{\citenamefont {Fournier}\ \emph {et~al.}(1998)\citenamefont
  {Fournier}, \citenamefont {Mohanty}, \citenamefont {Maiser}, \citenamefont
  {Darzens}, \citenamefont {Venkatesan}, \citenamefont {Lobb}, \citenamefont
  {Czjzek}, \citenamefont {Webb},\ and\ \citenamefont
  {Greene}}]{Fournier1998insulator}%
  \BibitemOpen
  \bibfield  {author} {\bibinfo {author} {\bibfnamefont {P.}~\bibnamefont
  {Fournier}}, \bibinfo {author} {\bibfnamefont {P.}~\bibnamefont {Mohanty}},
  \bibinfo {author} {\bibfnamefont {E.}~\bibnamefont {Maiser}}, \bibinfo
  {author} {\bibfnamefont {S.}~\bibnamefont {Darzens}}, \bibinfo {author}
  {\bibfnamefont {T.}~\bibnamefont {Venkatesan}}, \bibinfo {author}
  {\bibfnamefont {C.~J.}\ \bibnamefont {Lobb}}, \bibinfo {author}
  {\bibfnamefont {G.}~\bibnamefont {Czjzek}}, \bibinfo {author} {\bibfnamefont
  {R.~A.}\ \bibnamefont {Webb}}, \ and\ \bibinfo {author} {\bibfnamefont
  {R.~L.}\ \bibnamefont {Greene}},\ }\href {\doibase
  10.1103/PhysRevLett.81.4720} {\bibfield  {journal} {\bibinfo  {journal}
  {Phys. Rev. Lett.}\ }\textbf {\bibinfo {volume} {81}},\ \bibinfo {pages}
  {4720} (\bibinfo {year} {1998})}\BibitemShut {NoStop}%
\bibitem [{\citenamefont {Ono}\ \emph {et~al.}(2000)\citenamefont {Ono},
  \citenamefont {Ando}, \citenamefont {Murayama}, \citenamefont {Balakirev},
  \citenamefont {Betts},\ and\ \citenamefont {Boebinger}}]{Ono2000metal}%
  \BibitemOpen
  \bibfield  {author} {\bibinfo {author} {\bibfnamefont {S.}~\bibnamefont
  {Ono}}, \bibinfo {author} {\bibfnamefont {Y.}~\bibnamefont {Ando}}, \bibinfo
  {author} {\bibfnamefont {T.}~\bibnamefont {Murayama}}, \bibinfo {author}
  {\bibfnamefont {F.~F.}\ \bibnamefont {Balakirev}}, \bibinfo {author}
  {\bibfnamefont {J.~B.}\ \bibnamefont {Betts}}, \ and\ \bibinfo {author}
  {\bibfnamefont {G.~S.}\ \bibnamefont {Boebinger}},\ }\href {\doibase
  10.1103/PhysRevLett.85.638} {\bibfield  {journal} {\bibinfo  {journal} {Phys.
  Rev. Lett.}\ }\textbf {\bibinfo {volume} {85}},\ \bibinfo {pages} {638}
  (\bibinfo {year} {2000})}\BibitemShut {NoStop}%
\bibitem [{\citenamefont {Sekitani}\ \emph {et~al.}(2003)\citenamefont
  {Sekitani}, \citenamefont {Naito},\ and\ \citenamefont
  {Miura}}]{Sekitani2003Kondo}%
  \BibitemOpen
  \bibfield  {author} {\bibinfo {author} {\bibfnamefont {T.}~\bibnamefont
  {Sekitani}}, \bibinfo {author} {\bibfnamefont {M.}~\bibnamefont {Naito}}, \
  and\ \bibinfo {author} {\bibfnamefont {N.}~\bibnamefont {Miura}},\ }\href
  {\doibase 10.1103/PhysRevB.67.174503} {\bibfield  {journal} {\bibinfo
  {journal} {Phys. Rev. B}\ }\textbf {\bibinfo {volume} {67}},\ \bibinfo
  {pages} {174503} (\bibinfo {year} {2003})}\BibitemShut {NoStop}%
\bibitem [{\citenamefont {Wang}\ \emph {et~al.}(2005)\citenamefont {Wang},
  \citenamefont {Chen}, \citenamefont {Huang}, \citenamefont {Wang},
  \citenamefont {Xiong},\ and\ \citenamefont {Luo}}]{wang2005transport}%
  \BibitemOpen
  \bibfield  {author} {\bibinfo {author} {\bibfnamefont {C.}~\bibnamefont
  {Wang}}, \bibinfo {author} {\bibfnamefont {X.}~\bibnamefont {Chen}}, \bibinfo
  {author} {\bibfnamefont {L.}~\bibnamefont {Huang}}, \bibinfo {author}
  {\bibfnamefont {L.}~\bibnamefont {Wang}}, \bibinfo {author} {\bibfnamefont
  {Y.}~\bibnamefont {Xiong}}, \ and\ \bibinfo {author} {\bibfnamefont
  {X.}~\bibnamefont {Luo}},\ }\href {\doibase 10.1088/0953-8984/17/7/006}
  {\bibfield  {journal} {\bibinfo  {journal} {J. Phys.: Conden. Matt.}\
  }\textbf {\bibinfo {volume} {17}},\ \bibinfo {pages} {1127} (\bibinfo {year}
  {2005})}\BibitemShut {NoStop}%
\bibitem [{\citenamefont {Li}\ \emph {et~al.}(2016)\citenamefont {Li},
  \citenamefont {Tabis}, \citenamefont {Yu}, \citenamefont {Bari\ifmmode
  \check{s}\else \v{s}\fi{}i\ifmmode~\acute{c}\else \'{c}\fi{}},\ and\
  \citenamefont {Greven}}]{Li2016hidden}%
  \BibitemOpen
  \bibfield  {author} {\bibinfo {author} {\bibfnamefont {Y.}~\bibnamefont
  {Li}}, \bibinfo {author} {\bibfnamefont {W.}~\bibnamefont {Tabis}}, \bibinfo
  {author} {\bibfnamefont {G.}~\bibnamefont {Yu}}, \bibinfo {author}
  {\bibfnamefont {N.}~\bibnamefont {Bari\ifmmode \check{s}\else
  \v{s}\fi{}i\ifmmode~\acute{c}\else \'{c}\fi{}}}, \ and\ \bibinfo {author}
  {\bibfnamefont {M.}~\bibnamefont {Greven}},\ }\href {\doibase
  10.1103/PhysRevLett.117.197001} {\bibfield  {journal} {\bibinfo  {journal}
  {Phys. Rev. Lett.}\ }\textbf {\bibinfo {volume} {117}},\ \bibinfo {pages}
  {197001} (\bibinfo {year} {2016})}\BibitemShut {NoStop}%
\bibitem [{\citenamefont {Bari{\v{s}}i{\'c}}\ \emph {et~al.}(2019)\citenamefont
  {Bari{\v{s}}i{\'c}}, \citenamefont {Chan}, \citenamefont {Veit},
  \citenamefont {Dorow}, \citenamefont {Ge}, \citenamefont {Li}, \citenamefont
  {Tabis}, \citenamefont {Tang}, \citenamefont {Yu}, \citenamefont {Zhao},\
  and\ \citenamefont {Greven}}]{barivsic2019evidence}%
  \BibitemOpen
  \bibfield  {author} {\bibinfo {author} {\bibfnamefont {N.}~\bibnamefont
  {Bari{\v{s}}i{\'c}}}, \bibinfo {author} {\bibfnamefont {M.}~\bibnamefont
  {Chan}}, \bibinfo {author} {\bibfnamefont {M.}~\bibnamefont {Veit}}, \bibinfo
  {author} {\bibfnamefont {C.}~\bibnamefont {Dorow}}, \bibinfo {author}
  {\bibfnamefont {Y.}~\bibnamefont {Ge}}, \bibinfo {author} {\bibfnamefont
  {Y.}~\bibnamefont {Li}}, \bibinfo {author} {\bibfnamefont {W.}~\bibnamefont
  {Tabis}}, \bibinfo {author} {\bibfnamefont {Y.}~\bibnamefont {Tang}},
  \bibinfo {author} {\bibfnamefont {G.}~\bibnamefont {Yu}}, \bibinfo {author}
  {\bibfnamefont {X.}~\bibnamefont {Zhao}}, \ and\ \bibinfo {author}
  {\bibfnamefont {M.}~\bibnamefont {Greven}},\ }\href {\doibase
  10.1088/1367-2630/ab4d0f} {\bibfield  {journal} {\bibinfo  {journal} {New J.
  Phys.}\ }\textbf {\bibinfo {volume} {21}},\ \bibinfo {pages} {113007}
  (\bibinfo {year} {2019})}\BibitemShut {NoStop}%
\bibitem [{\citenamefont {Pelc}\ \emph {et~al.}(2019)\citenamefont {Pelc},
  \citenamefont {Pop{\v{c}}evi{\'c}}, \citenamefont {Po{\v{z}}ek},
  \citenamefont {Greven},\ and\ \citenamefont
  {Bari{\v{s}}i{\'c}}}]{pelc2019unusual}%
  \BibitemOpen
  \bibfield  {author} {\bibinfo {author} {\bibfnamefont {D.}~\bibnamefont
  {Pelc}}, \bibinfo {author} {\bibfnamefont {P.}~\bibnamefont
  {Pop{\v{c}}evi{\'c}}}, \bibinfo {author} {\bibfnamefont {M.}~\bibnamefont
  {Po{\v{z}}ek}}, \bibinfo {author} {\bibfnamefont {M.}~\bibnamefont {Greven}},
  \ and\ \bibinfo {author} {\bibfnamefont {N.}~\bibnamefont
  {Bari{\v{s}}i{\'c}}},\ }\href {\doibase 10.1126/sciadv.aau4538} {\bibfield
  {journal} {\bibinfo  {journal} {Sci. Adv.}\ }\textbf {\bibinfo {volume}
  {5}},\ \bibinfo {pages} {eaau4538} (\bibinfo {year} {2019})}\BibitemShut
  {NoStop}%
\bibitem [{\citenamefont {Lee}\ and\ \citenamefont
  {Ramakrishnan}(1985)}]{Lee1985disordered}%
  \BibitemOpen
  \bibfield  {author} {\bibinfo {author} {\bibfnamefont {P.~A.}\ \bibnamefont
  {Lee}}\ and\ \bibinfo {author} {\bibfnamefont {T.~V.}\ \bibnamefont
  {Ramakrishnan}},\ }\href {\doibase 10.1103/RevModPhys.57.287} {\bibfield
  {journal} {\bibinfo  {journal} {Rev. Mod. Phys.}\ }\textbf {\bibinfo {volume}
  {57}},\ \bibinfo {pages} {287} (\bibinfo {year} {1985})}\BibitemShut
  {NoStop}%
\bibitem [{\citenamefont {Beloborodov}\ \emph {et~al.}(2003)\citenamefont
  {Beloborodov}, \citenamefont {Efetov}, \citenamefont {Lopatin},\ and\
  \citenamefont {Vinokur}}]{Beloborodov2003transport}%
  \BibitemOpen
  \bibfield  {author} {\bibinfo {author} {\bibfnamefont {I.~S.}\ \bibnamefont
  {Beloborodov}}, \bibinfo {author} {\bibfnamefont {K.~B.}\ \bibnamefont
  {Efetov}}, \bibinfo {author} {\bibfnamefont {A.~V.}\ \bibnamefont {Lopatin}},
  \ and\ \bibinfo {author} {\bibfnamefont {V.~M.}\ \bibnamefont {Vinokur}},\
  }\href {\doibase 10.1103/PhysRevLett.91.246801} {\bibfield  {journal}
  {\bibinfo  {journal} {Phys. Rev. Lett.}\ }\textbf {\bibinfo {volume} {91}},\
  \bibinfo {pages} {246801} (\bibinfo {year} {2003})}\BibitemShut {NoStop}%
\bibitem [{\citenamefont {Zhou}\ \emph {et~al.}(2018)\citenamefont {Zhou},
  \citenamefont {Peets}, \citenamefont {Morgan}, \citenamefont {Huttema},
  \citenamefont {Murphy}, \citenamefont {Thewalt}, \citenamefont {Truncik},
  \citenamefont {Turner}, \citenamefont {Koenig}, \citenamefont {Waldram},
  \citenamefont {Hosseini}, \citenamefont {Liang}, \citenamefont {Bonn},
  \citenamefont {Hardy},\ and\ \citenamefont {Broun}}]{Zhou2018logarithmic}%
  \BibitemOpen
  \bibfield  {author} {\bibinfo {author} {\bibfnamefont {X.~Q.}\ \bibnamefont
  {Zhou}}, \bibinfo {author} {\bibfnamefont {D.~C.}\ \bibnamefont {Peets}},
  \bibinfo {author} {\bibfnamefont {B.}~\bibnamefont {Morgan}}, \bibinfo
  {author} {\bibfnamefont {W.~A.}\ \bibnamefont {Huttema}}, \bibinfo {author}
  {\bibfnamefont {N.~C.}\ \bibnamefont {Murphy}}, \bibinfo {author}
  {\bibfnamefont {E.}~\bibnamefont {Thewalt}}, \bibinfo {author} {\bibfnamefont
  {C.~J.~S.}\ \bibnamefont {Truncik}}, \bibinfo {author} {\bibfnamefont
  {P.~J.}\ \bibnamefont {Turner}}, \bibinfo {author} {\bibfnamefont {A.~J.}\
  \bibnamefont {Koenig}}, \bibinfo {author} {\bibfnamefont {J.~R.}\
  \bibnamefont {Waldram}}, \bibinfo {author} {\bibfnamefont {A.}~\bibnamefont
  {Hosseini}}, \bibinfo {author} {\bibfnamefont {R.}~\bibnamefont {Liang}},
  \bibinfo {author} {\bibfnamefont {D.~A.}\ \bibnamefont {Bonn}}, \bibinfo
  {author} {\bibfnamefont {W.~N.}\ \bibnamefont {Hardy}}, \ and\ \bibinfo
  {author} {\bibfnamefont {D.~M.}\ \bibnamefont {Broun}},\ }\href {\doibase
  10.1103/PhysRevLett.121.267004} {\bibfield  {journal} {\bibinfo  {journal}
  {Phys. Rev. Lett.}\ }\textbf {\bibinfo {volume} {121}},\ \bibinfo {pages}
  {267004} (\bibinfo {year} {2018})}\BibitemShut {NoStop}%
\bibitem [{\citenamefont {Beloborodov}\ \emph {et~al.}(2007)\citenamefont
  {Beloborodov}, \citenamefont {Lopatin}, \citenamefont {Vinokur},\ and\
  \citenamefont {Efetov}}]{Beloborodov2007grnular}%
  \BibitemOpen
  \bibfield  {author} {\bibinfo {author} {\bibfnamefont {I.~S.}\ \bibnamefont
  {Beloborodov}}, \bibinfo {author} {\bibfnamefont {A.~V.}\ \bibnamefont
  {Lopatin}}, \bibinfo {author} {\bibfnamefont {V.~M.}\ \bibnamefont
  {Vinokur}}, \ and\ \bibinfo {author} {\bibfnamefont {K.~B.}\ \bibnamefont
  {Efetov}},\ }\href {\doibase 10.1103/RevModPhys.79.469} {\bibfield  {journal}
  {\bibinfo  {journal} {Rev. Mod. Phys.}\ }\textbf {\bibinfo {volume} {79}},\
  \bibinfo {pages} {469} (\bibinfo {year} {2007})}\BibitemShut {NoStop}%
\bibitem [{\citenamefont {Efetov}\ and\ \citenamefont
  {Tschersich}(2003)}]{Efetov2003coulomb}%
  \BibitemOpen
  \bibfield  {author} {\bibinfo {author} {\bibfnamefont {K.~B.}\ \bibnamefont
  {Efetov}}\ and\ \bibinfo {author} {\bibfnamefont {A.}~\bibnamefont
  {Tschersich}},\ }\href {\doibase 10.1103/PhysRevB.67.174205} {\bibfield
  {journal} {\bibinfo  {journal} {Phys. Rev. B}\ }\textbf {\bibinfo {volume}
  {67}},\ \bibinfo {pages} {174205} (\bibinfo {year} {2003})}\BibitemShut
  {NoStop}%
\bibitem [{\citenamefont {Sachser}\ \emph {et~al.}(2011)\citenamefont
  {Sachser}, \citenamefont {Porrati}, \citenamefont {Schwalb},\ and\
  \citenamefont {Huth}}]{Sachser2011universal}%
  \BibitemOpen
  \bibfield  {author} {\bibinfo {author} {\bibfnamefont {R.}~\bibnamefont
  {Sachser}}, \bibinfo {author} {\bibfnamefont {F.}~\bibnamefont {Porrati}},
  \bibinfo {author} {\bibfnamefont {C.~H.}\ \bibnamefont {Schwalb}}, \ and\
  \bibinfo {author} {\bibfnamefont {M.}~\bibnamefont {Huth}},\ }\href {\doibase
  10.1103/PhysRevLett.107.206803} {\bibfield  {journal} {\bibinfo  {journal}
  {Phys. Rev. Lett.}\ }\textbf {\bibinfo {volume} {107}},\ \bibinfo {pages}
  {206803} (\bibinfo {year} {2011})}\BibitemShut {NoStop}%
\bibitem [{\citenamefont {Pan}\ \emph {et~al.}(2001)\citenamefont {Pan},
  \citenamefont {O'neal}, \citenamefont {Badzey}, \citenamefont {Chamon},
  \citenamefont {Ding}, \citenamefont {Engelbrecht}, \citenamefont {Wang},
  \citenamefont {Eisaki}, \citenamefont {Uchida}, \citenamefont {Gupta},
  \citenamefont {Ng}, \citenamefont {Hudson}, \citenamefont {Lang},\ and\
  \citenamefont {Davis}}]{pan2001microscopic}%
  \BibitemOpen
  \bibfield  {author} {\bibinfo {author} {\bibfnamefont {S.~H.}\ \bibnamefont
  {Pan}}, \bibinfo {author} {\bibfnamefont {J.~P.}\ \bibnamefont {O'neal}},
  \bibinfo {author} {\bibfnamefont {R.~L.}\ \bibnamefont {Badzey}}, \bibinfo
  {author} {\bibfnamefont {C.}~\bibnamefont {Chamon}}, \bibinfo {author}
  {\bibfnamefont {H.}~\bibnamefont {Ding}}, \bibinfo {author} {\bibfnamefont
  {J.~R.}\ \bibnamefont {Engelbrecht}}, \bibinfo {author} {\bibfnamefont
  {Z.}~\bibnamefont {Wang}}, \bibinfo {author} {\bibfnamefont {H.}~\bibnamefont
  {Eisaki}}, \bibinfo {author} {\bibfnamefont {S.}~\bibnamefont {Uchida}},
  \bibinfo {author} {\bibfnamefont {A.~K.}\ \bibnamefont {Gupta}}, \bibinfo
  {author} {\bibfnamefont {K.~W.}\ \bibnamefont {Ng}}, \bibinfo {author}
  {\bibfnamefont {E.~W.}\ \bibnamefont {Hudson}}, \bibinfo {author}
  {\bibfnamefont {K.~M.}\ \bibnamefont {Lang}}, \ and\ \bibinfo {author}
  {\bibfnamefont {J.~C.}\ \bibnamefont {Davis}},\ }\href {\doibase
  10.1038/35095012} {\bibfield  {journal} {\bibinfo  {journal} {Nature}\
  }\textbf {\bibinfo {volume} {413}},\ \bibinfo {pages} {282} (\bibinfo {year}
  {2001})}\BibitemShut {NoStop}%
\bibitem [{\citenamefont {Parra}\ \emph {et~al.}(2021)\citenamefont {Parra},
  \citenamefont {Niestemski}, \citenamefont {Contryman}, \citenamefont
  {Giraldo~Gallo}, \citenamefont {Geballe}, \citenamefont {Fisher},\ and\
  \citenamefont {Manoharan}}]{parra2021signatures}%
  \BibitemOpen
  \bibfield  {author} {\bibinfo {author} {\bibfnamefont {C.}~\bibnamefont
  {Parra}}, \bibinfo {author} {\bibfnamefont {F.~C.}\ \bibnamefont
  {Niestemski}}, \bibinfo {author} {\bibfnamefont {A.~W.}\ \bibnamefont
  {Contryman}}, \bibinfo {author} {\bibfnamefont {P.}~\bibnamefont
  {Giraldo~Gallo}}, \bibinfo {author} {\bibfnamefont {T.~H.}\ \bibnamefont
  {Geballe}}, \bibinfo {author} {\bibfnamefont {I.~R.}\ \bibnamefont {Fisher}},
  \ and\ \bibinfo {author} {\bibfnamefont {H.~C.}\ \bibnamefont {Manoharan}},\
  }\href {\doibase 10.1073/pnas.2017810118} {\bibfield  {journal} {\bibinfo
  {journal} {Proc. Natl. Acad. Sci. USA}\ }\textbf {\bibinfo {volume} {118}},\
  \bibinfo {pages} {e2017810118} (\bibinfo {year} {2021})}\BibitemShut
  {NoStop}%
\bibitem [{\citenamefont {Pelc}\ \emph {et~al.}(2018)\citenamefont {Pelc},
  \citenamefont {Vu{\v{c}}kovi{\'c}}, \citenamefont {Grbi{\'c}}, \citenamefont
  {Po{\v{z}}ek}, \citenamefont {Yu}, \citenamefont {Sasagawa}, \citenamefont
  {Greven},\ and\ \citenamefont {Bari{\v{s}}i{\'c}}}]{pelc2018emergence}%
  \BibitemOpen
  \bibfield  {author} {\bibinfo {author} {\bibfnamefont {D.}~\bibnamefont
  {Pelc}}, \bibinfo {author} {\bibfnamefont {M.}~\bibnamefont
  {Vu{\v{c}}kovi{\'c}}}, \bibinfo {author} {\bibfnamefont {M.~S.}\ \bibnamefont
  {Grbi{\'c}}}, \bibinfo {author} {\bibfnamefont {M.}~\bibnamefont
  {Po{\v{z}}ek}}, \bibinfo {author} {\bibfnamefont {G.}~\bibnamefont {Yu}},
  \bibinfo {author} {\bibfnamefont {T.}~\bibnamefont {Sasagawa}}, \bibinfo
  {author} {\bibfnamefont {M.}~\bibnamefont {Greven}}, \ and\ \bibinfo {author}
  {\bibfnamefont {N.}~\bibnamefont {Bari{\v{s}}i{\'c}}},\ }\href {\doibase
  10.1038/s41467-018-06707-y} {\bibfield  {journal} {\bibinfo  {journal} {Nat.
  Commun.}\ }\textbf {\bibinfo {volume} {9}},\ \bibinfo {pages} {1} (\bibinfo
  {year} {2018})}\BibitemShut {NoStop}%
\bibitem [{\citenamefont {Pop{\v{c}}evi{\'c}}\ \emph
  {et~al.}(2018)\citenamefont {Pop{\v{c}}evi{\'c}}, \citenamefont {Pelc},
  \citenamefont {Tang}, \citenamefont {Velebit}, \citenamefont {Anderson},
  \citenamefont {Nagarajan}, \citenamefont {Yu}, \citenamefont {Po{\v{z}}ek},
  \citenamefont {Bari{\v{s}}i{\'c}},\ and\ \citenamefont
  {Greven}}]{popvcevic2018percolative}%
  \BibitemOpen
  \bibfield  {author} {\bibinfo {author} {\bibfnamefont {P.}~\bibnamefont
  {Pop{\v{c}}evi{\'c}}}, \bibinfo {author} {\bibfnamefont {D.}~\bibnamefont
  {Pelc}}, \bibinfo {author} {\bibfnamefont {Y.}~\bibnamefont {Tang}}, \bibinfo
  {author} {\bibfnamefont {K.}~\bibnamefont {Velebit}}, \bibinfo {author}
  {\bibfnamefont {Z.}~\bibnamefont {Anderson}}, \bibinfo {author}
  {\bibfnamefont {V.}~\bibnamefont {Nagarajan}}, \bibinfo {author}
  {\bibfnamefont {G.}~\bibnamefont {Yu}}, \bibinfo {author} {\bibfnamefont
  {M.}~\bibnamefont {Po{\v{z}}ek}}, \bibinfo {author} {\bibfnamefont
  {N.}~\bibnamefont {Bari{\v{s}}i{\'c}}}, \ and\ \bibinfo {author}
  {\bibfnamefont {M.}~\bibnamefont {Greven}},\ }\href {\doibase
  10.1038/s41535-018-0115-2} {\bibfield  {journal} {\bibinfo  {journal} {npj
  Quan. Mater.}\ }\textbf {\bibinfo {volume} {3}},\ \bibinfo {pages} {1}
  (\bibinfo {year} {2018})}\BibitemShut {NoStop}%
\bibitem [{\citenamefont {Yu}\ \emph {et~al.}(2019)\citenamefont {Yu},
  \citenamefont {Xia}, \citenamefont {Pelc}, \citenamefont {He}, \citenamefont
  {Kaneko}, \citenamefont {Sasagawa}, \citenamefont {Li}, \citenamefont {Zhao},
  \citenamefont {Bari\ifmmode \check{s}\else \v{s}\fi{}i\ifmmode~\acute{c}\else
  \'{c}\fi{}}, \citenamefont {Shekhter},\ and\ \citenamefont
  {Greven}}]{Yu2019universal}%
  \BibitemOpen
  \bibfield  {author} {\bibinfo {author} {\bibfnamefont {G.}~\bibnamefont
  {Yu}}, \bibinfo {author} {\bibfnamefont {D.~D.}\ \bibnamefont {Xia}},
  \bibinfo {author} {\bibfnamefont {D.}~\bibnamefont {Pelc}}, \bibinfo {author}
  {\bibfnamefont {R.~H.}\ \bibnamefont {He}}, \bibinfo {author} {\bibfnamefont
  {N.~H.}\ \bibnamefont {Kaneko}}, \bibinfo {author} {\bibfnamefont
  {T.}~\bibnamefont {Sasagawa}}, \bibinfo {author} {\bibfnamefont
  {Y.}~\bibnamefont {Li}}, \bibinfo {author} {\bibfnamefont {X.}~\bibnamefont
  {Zhao}}, \bibinfo {author} {\bibfnamefont {N.}~\bibnamefont {Bari\ifmmode
  \check{s}\else \v{s}\fi{}i\ifmmode~\acute{c}\else \'{c}\fi{}}}, \bibinfo
  {author} {\bibfnamefont {A.}~\bibnamefont {Shekhter}}, \ and\ \bibinfo
  {author} {\bibfnamefont {M.}~\bibnamefont {Greven}},\ }\href {\doibase
  10.1103/PhysRevB.99.214502} {\bibfield  {journal} {\bibinfo  {journal} {Phys.
  Rev. B}\ }\textbf {\bibinfo {volume} {99}},\ \bibinfo {pages} {214502}
  (\bibinfo {year} {2019})}\BibitemShut {NoStop}%
\bibitem [{\citenamefont {Tajima}\ \emph {et~al.}(1990)\citenamefont {Tajima},
  \citenamefont {Uchida}, \citenamefont {Ishibashi}, \citenamefont {Ido},
  \citenamefont {Takagi}, \citenamefont {Arima},\ and\ \citenamefont
  {Tokura}}]{tajima1990optical}%
  \BibitemOpen
  \bibfield  {author} {\bibinfo {author} {\bibfnamefont {S.}~\bibnamefont
  {Tajima}}, \bibinfo {author} {\bibfnamefont {S.}~\bibnamefont {Uchida}},
  \bibinfo {author} {\bibfnamefont {S.}~\bibnamefont {Ishibashi}}, \bibinfo
  {author} {\bibfnamefont {T.}~\bibnamefont {Ido}}, \bibinfo {author}
  {\bibfnamefont {H.}~\bibnamefont {Takagi}}, \bibinfo {author} {\bibfnamefont
  {T.}~\bibnamefont {Arima}}, \ and\ \bibinfo {author} {\bibfnamefont
  {Y.}~\bibnamefont {Tokura}},\ }\href {\doibase 10.1016/0921-4534(90)90113-S}
  {\bibfield  {journal} {\bibinfo  {journal} {Physica C}\ }\textbf {\bibinfo
  {volume} {168}},\ \bibinfo {pages} {117} (\bibinfo {year}
  {1990})}\BibitemShut {NoStop}%
\bibitem [{\citenamefont {Tajima}\ \emph {et~al.}(1991)\citenamefont {Tajima},
  \citenamefont {Ido}, \citenamefont {Ishibashi}, \citenamefont {Itoh},
  \citenamefont {Eisaki}, \citenamefont {Mizuo}, \citenamefont {Arima},
  \citenamefont {Takagi},\ and\ \citenamefont {Uchida}}]{Tajima1991optical}%
  \BibitemOpen
  \bibfield  {author} {\bibinfo {author} {\bibfnamefont {S.}~\bibnamefont
  {Tajima}}, \bibinfo {author} {\bibfnamefont {T.}~\bibnamefont {Ido}},
  \bibinfo {author} {\bibfnamefont {S.}~\bibnamefont {Ishibashi}}, \bibinfo
  {author} {\bibfnamefont {T.}~\bibnamefont {Itoh}}, \bibinfo {author}
  {\bibfnamefont {H.}~\bibnamefont {Eisaki}}, \bibinfo {author} {\bibfnamefont
  {Y.}~\bibnamefont {Mizuo}}, \bibinfo {author} {\bibfnamefont
  {T.}~\bibnamefont {Arima}}, \bibinfo {author} {\bibfnamefont
  {H.}~\bibnamefont {Takagi}}, \ and\ \bibinfo {author} {\bibfnamefont
  {S.}~\bibnamefont {Uchida}},\ }\href {\doibase 10.1103/PhysRevB.43.10496}
  {\bibfield  {journal} {\bibinfo  {journal} {Phys. Rev. B}\ }\textbf {\bibinfo
  {volume} {43}},\ \bibinfo {pages} {10496} (\bibinfo {year}
  {1991})}\BibitemShut {NoStop}%
\bibitem [{\citenamefont {Wang}\ \emph {et~al.}(2020)\citenamefont {Wang},
  \citenamefont {Guan}, \citenamefont {Xiong}, \citenamefont {Zhang},
  \citenamefont {Fan}, \citenamefont {Zhu}, \citenamefont {Song}, \citenamefont
  {Ma},\ and\ \citenamefont {Xue}}]{Wang2020electronic}%
  \BibitemOpen
  \bibfield  {author} {\bibinfo {author} {\bibfnamefont {R.~F.}\ \bibnamefont
  {Wang}}, \bibinfo {author} {\bibfnamefont {J.}~\bibnamefont {Guan}}, \bibinfo
  {author} {\bibfnamefont {Y.~L.}\ \bibnamefont {Xiong}}, \bibinfo {author}
  {\bibfnamefont {X.~F.}\ \bibnamefont {Zhang}}, \bibinfo {author}
  {\bibfnamefont {J.~Q.}\ \bibnamefont {Fan}}, \bibinfo {author} {\bibfnamefont
  {J.}~\bibnamefont {Zhu}}, \bibinfo {author} {\bibfnamefont {C.~L.}\
  \bibnamefont {Song}}, \bibinfo {author} {\bibfnamefont {X.~C.}\ \bibnamefont
  {Ma}}, \ and\ \bibinfo {author} {\bibfnamefont {Q.~K.}\ \bibnamefont {Xue}},\
  }\href {\doibase 10.1103/PhysRevB.102.100508} {\bibfield  {journal} {\bibinfo
   {journal} {Phys. Rev. B}\ }\textbf {\bibinfo {volume} {102}},\ \bibinfo
  {pages} {100508} (\bibinfo {year} {2020})}\BibitemShut {NoStop}%
\end{thebibliography}
%

%%%%%%%%%% Merge with supplemental materials %%%%%%%%%%
% \pagebreak
\widetext
\newpage
%%%%%%%%%% Merge with supplemental materials %%%%%%%%%%
%%%%%%%%%% Prefix a "S" to all equations, figures, tables and reset the counter %%%%%%%%%%
\setcounter{equation}{0}
\setcounter{figure}{0}
\setcounter{table}{0}
\setcounter{page}{5}
\makeatletter
\renewcommand{\theequation}{S\arabic{equation}}
\renewcommand{\thefigure}{S\arabic{figure}}
\renewcommand{\bibnumfmt}[1]{[S#1]}
\renewcommand{\citenumfont}[1]{S#1}
%%%%%%%%%% Prefix a "S" to all equations, figures, tables and reset the counter %%%%%%%%%%

\onecolumngrid
\Large
{\textbf{Supplemental Material for:} }
\begin{center}
\textbf{\large Percolative Superconductivity in Electron-Doped Sr$_{1-x}$Eu$_{x}$CuO$_{2+y}$ Films}
\end{center}

\small
\maketitle
\onecolumngrid

This supplement includes:

Figs.\ S1 to S9 and Captions
\\
\\

\begin{figure}[h]
\includegraphics[width=0.52\columnwidth]{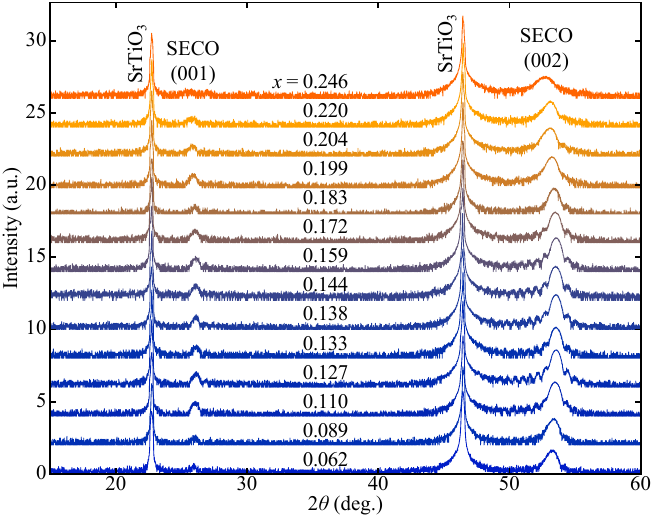}
\caption{XRD spectra for SECO epitaxial thin films ($\sim$ 50 unit cells) over an extended scattering angle 2$\theta$ from 15$^o$ to 60$^o$. The Eu doping content ranges from $x$ = 0.062 to 0.246, all of which exhibit no other phase except for the electron-doped infinite-layer cuprate phase. The sharp Bragg  peaks and Kiessig fringes surround the main (002) Bragg peak demonstrate the good crystallinity of MBE-grown SECO epitaxial films.
}
\end{figure}

\begin{figure}[h]
\includegraphics[width=0.52\columnwidth]{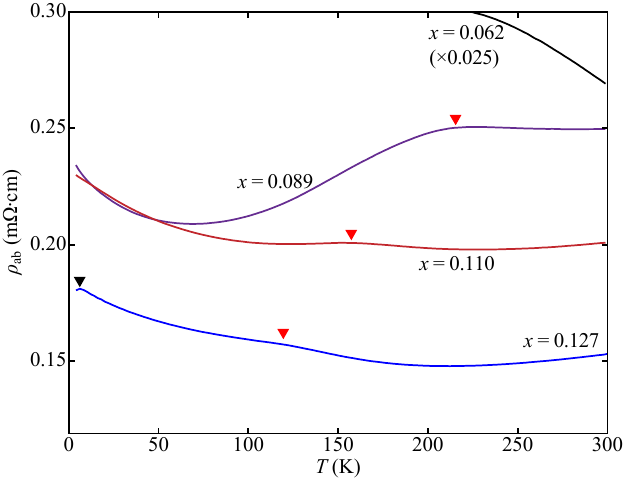}
\caption{Temperature dependence of the in-plane electrical resistivity $\rho_{\textrm{ab}}$ in underdoped SECO epitaxial films. At $x$ = 0.089, the resistivity $\rho_{\textrm{ab}}$ shows a faint insulating-like behavior ($\rho_{\textrm{ab}}$/d$T$ $<$ 0) at high temperatures, begins to rapidly decrease at a characteristic temperature signaled by the red triangle and then a pronounced upturn toward zero temperature, forming a hump feature at $T_{\textrm{hump}}$ $\sim$ 210 K. The resistivity $\rho_{\textrm{ab}}$ for $x >$ 0.089 shows a barely metallic behavior ($\rho_{\textrm{ab}}$/d$T$ $>$0) at high temperatures, but the hump feature remains and Thump reduces with $x$. Note that a small drop, marked by the black triangle, is caused by the onset of superconductivity for the $x$ = 0.127 sample.
}
\end{figure}

\begin{figure}[h]
\includegraphics[width=1\columnwidth]{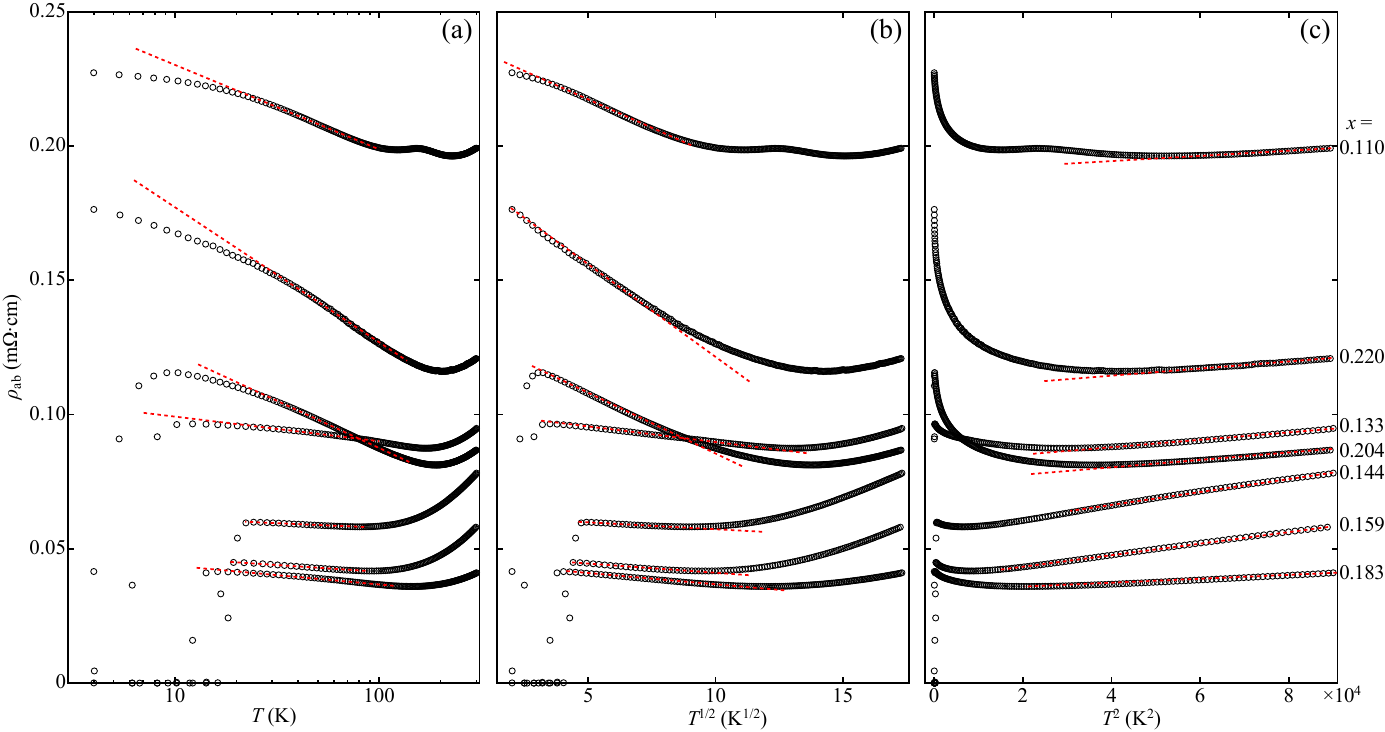}
\caption{(color online) (a-c) In-plane electrical resistivity $\rho_{\textrm{ab}}$ (empty circles) plotted against the (a) logarithm, (b) square-root and (c) square of temperature $T$, respectively, in seven SECO films at varied $x$. The straight dashed lines emphasize the logarithmic (log(1/$T$)) or square-root ($\sqrt{T}$) at low $T$ and quadratic ($T^2$) behaviors at high $T$ for the resistivity $\rho_{\textrm{ab}}$.
}
\end{figure}

\begin{figure}[h]
\includegraphics[width=0.65\columnwidth]{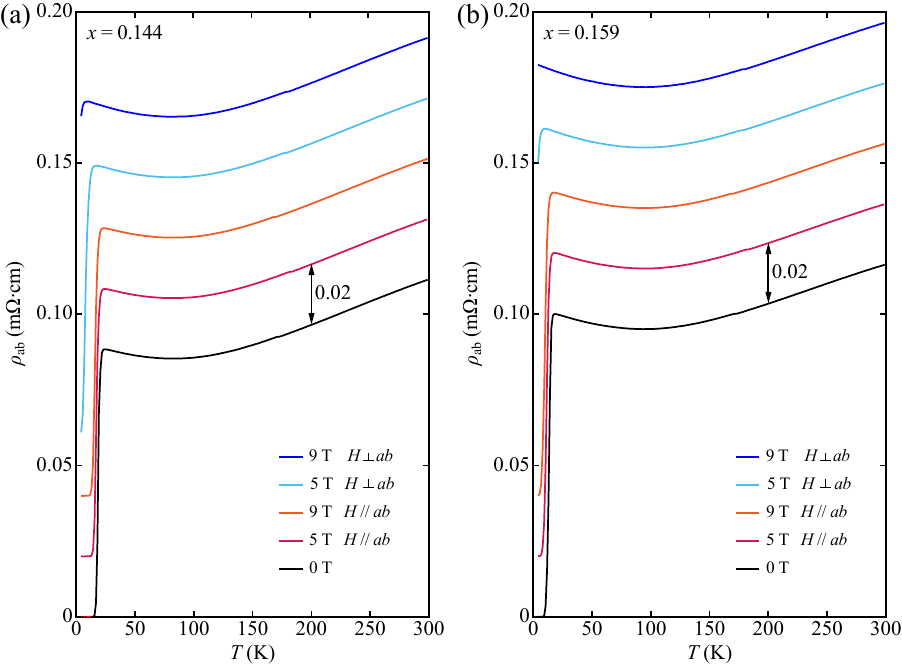}
\caption{(color online) (a, b) In-plane electrical resistivity $\rho_{\textrm{ab}}$ versus $T$ under various magnetic fields in the $x$ = 0.144 and $x$ = 0.159 samples, respectively. For clarity the curves under magnetic fields have been offset by 0.02 m$\Omega\cdot$cm. The square-root low-T upturn in the normal-state $\rho_{\textrm{ab}}$ is essentially unchanged against both in-plane and out-of-plane magnetic fields.
}
\end{figure}

\begin{figure}[h]
\includegraphics[width=0.81\columnwidth]{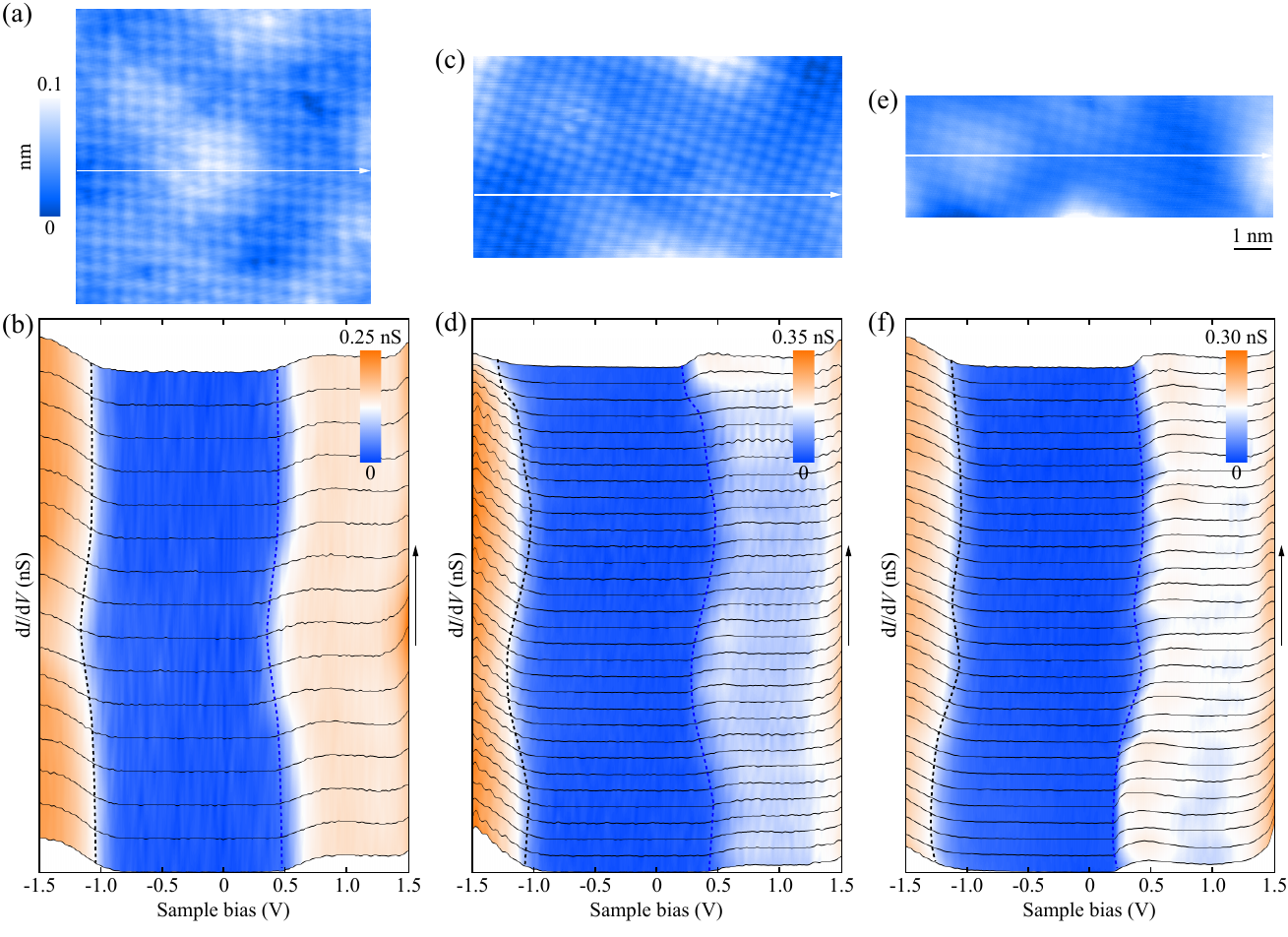}
\caption{(color online) (a-f) Universality of the nanoscale phase separation between bright and dark domains, which are characteristic of electron-rich and electron-poor electronic structures, respectively. The tunneling spectra in the bottom panel were measured along the corresponding white arrows of the STM topographies in the top panel, with the blue and black dashes marking the UHB and CTB onsets, respectively. Clearly, the $E_\textrm{F}$ is closer to UHB in the topographically bright regions, meaning more electron doping there. The tunneling junction was stabilized at $V$ = -1.5 V and $I$ = 100 pA, while the STM topographic images were acquired at $V$ = -1.0 V and $I$ = 20 pA except for (a) $V$ = 0.8 V. (a, b) $x$ = 0.064; (c, d) $x$ = 0.109; (c, d) $x$ = 0.118.
}
\end{figure}

\begin{figure}[h]
\includegraphics[width=0.35\columnwidth]{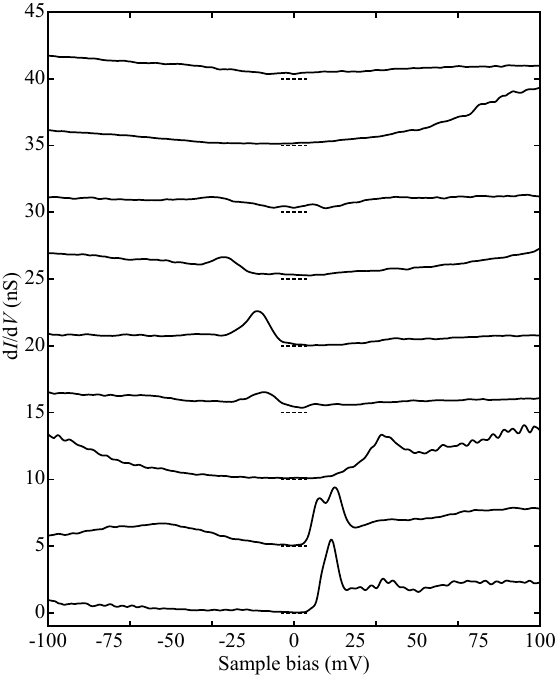}
\caption{(color online) A series of tunneling d$I$/d$V$ spectra ($V$ = -1.5 V and $I$ = 100 pA) taken in various non-superconducting regions from four SECO samples (0.114 $<x<$ 0.164). The spectra have been vertically offset for clarity, with their zero conductance positions marked by the short horizontal dashes. The finite density of states around the Fermi level are characteristic of the metallic-like electronic structure in these non-superconducting regions.
}
\end{figure}

\begin{figure}[h]
\includegraphics[width=0.5\columnwidth]{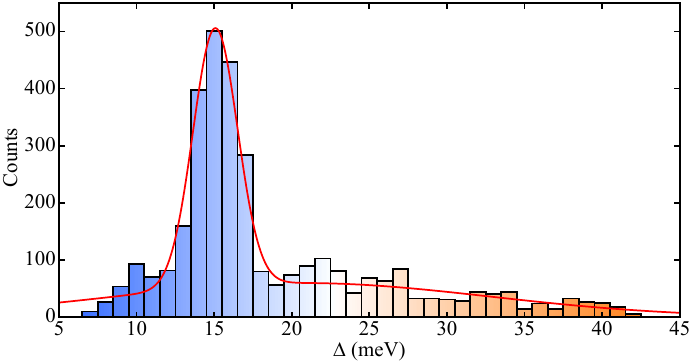}
\caption{(color online) Histogram of the superconducting gap $\Delta$ from various regions in twelve SECO samples, showing a pronounced peak around 15 meV. The red curve is a guide to the eyes.
}
\end{figure}

\begin{figure}[h]
\includegraphics[width=0.5\columnwidth]{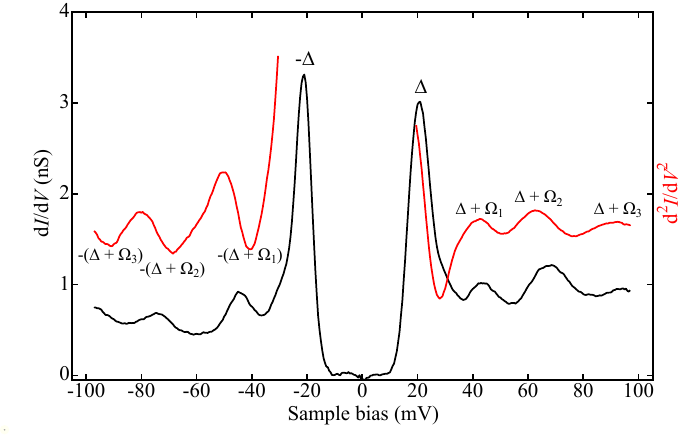}
\caption{(color online) One representative superconducting spectrum (setpoint: $V$ = -100 mV and $I$ = 100 pA) and its derivative, from which $\Omega_{1,2,3}$ can be extracted. The phonon modes exhibit themselves, via a strong coupling to the paired electrons, in the low-lying quasiparticle states at energy $E$ = $\Delta$ + $\Omega$ ($\Omega$ is the phonon energy).
}
\end{figure}

\begin{figure}[h]
\includegraphics[width=0.78\columnwidth]{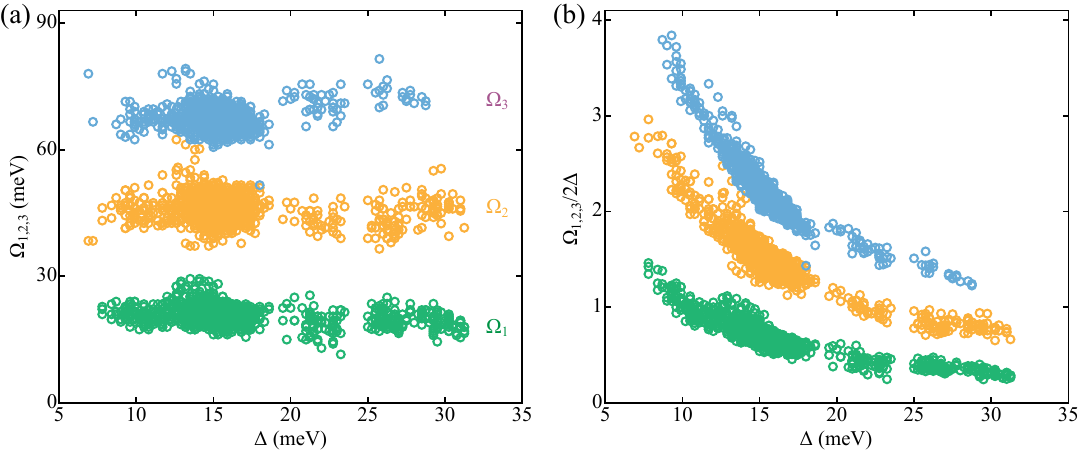}
\caption{(color online)  (a) Dependence of the three phonon energies $\Omega_{1,2,3}$ on the local superconducting gap $\Delta$, acquired at 4.8 K. (b) $\Omega_{1,2,3}$/2$\Delta$ plotted as a function of $\Delta$. Notice that the $\Omega_{1,2,3}$/2$\Delta$ exceeds unity at smaller $\Delta$ for $\Omega_{1,2}$ and all $\Delta$ for $\Omega_{3}$.
}
\end{figure}

\end{document}